\documentclass[journal=jacsat,manuscript=article]{achemso}
\setkeys{acs}{keywords = true}
\setkeys{acs}{abbreviations= true}
\usepackage{chemformula} % Formula subscripts using \ch{}
\usepackage[T1]{fontenc} % Use modern font encodings
\usepackage{xcolor}
\usepackage{graphicx}
\graphicspath{{figures/}}
\usepackage{csquotes}
\usepackage{subcaption}
\usepackage{lineno}
\usepackage{colortbl}
\usepackage{rotating}
\usepackage{adjustbox}
\usepackage{tikz}
\usetikzlibrary{arrows}
\usetikzlibrary{positioning}
\usetikzlibrary{shapes.geometric}
\usetikzlibrary{matrix, positioning, fit}
\usepackage[inline]{enumitem}
\usepackage{blindtext}
\usepackage{forest}
\tikzset{
Above/.style={
  midway,
  above,
  font=\scriptsize,
  text width=1.5cm,
  align=center,
  },
Below/.style={
  midway,
  below,
  font=\scriptsize,
  text width=1.5cm,
  align=center
  }
}
\tikzset{every label/.style={xshift=6ex, yshift=-12ex, text width=10ex, align=right, inner sep=1pt, font=\scriptsize, text=black}}
\usetikzlibrary{calc}
\tikzset{
    ncbar angle/.initial=90,
    ncbar/.style={
        to path=(\tikztostart)
        -- ($(\tikztostart)!#1!\pgfkeysvalueof{/tikz/ncbar angle}:(\tikztotarget)$)
        -- ($(\tikztotarget)!($(\tikztostart)!#1!\pgfkeysvalueof{/tikz/ncbar angle}:(\tikztotarget)$)!\pgfkeysvalueof{/tikz/ncbar angle}:(\tikztostart)$)
        -- (\tikztotarget)
    },
    ncbar/.default=0.5cm,
}

\tikzset{square left brace/.style={ncbar=0.5cm}}
\tikzset{square right brace/.style={ncbar=-0.5cm}}
\tikzset{round left paren/.style={ncbar=0.5cm,out=120,in=-120}}
\tikzset{round right paren/.style={ncbar=0.5cm,out=60,in=-60}}
\newcommand\longvdots[1]{\raisebox{1em}{\rotatebox{-90}{\hbox to #1 {\dotfill}}}}
\usepackage{hyperref}
\hypersetup{
  colorlinks,
  citecolor=violet,
  linkcolor=red,
  urlcolor=blue}
\usepackage{titlesec}

\titleclass{\subsubsubsection}{straight}[\subsection]
\newcounter{subsubsubsection}[subsubsection]
\renewcommand\thesubsubsubsection{\thesubsubsection.\arabic{subsubsubsection}}
 
\titleformat{\subsubsubsection}
  {\normalfont\normalsize\bfseries}{\thesubsubsubsection}{1em}{}
\titlespacing*{\subsubsubsection}
{0pt}{3.25ex plus 1ex minus .2ex}{1.5ex plus .2ex}
\newcommand*{\ie}{i.e.}

\newcommand*{\eg}{e.g.}
\newcommand*{\fig }{Fig.} 
 
\newcommand*{\sect }{Sec.}

\newcommand*{\tab }{Table}

\newcommand*{\hsd}{Hofmeister dataset}
\newcommand*{\rf}{Random Forest}
\newcommand*{\tsf}{\textit{tsfresh}}
\newcommand*{\laz}{\textit{LazyPredictor}}
\author{Jeyashree Krishnan}
\affiliation{Joint Research Center for Computational Biomedicine, RWTH Aachen University, Germany}
\email{krishnan@aices.rwth-aachen.de}
\author{Zeyu Lian}
\affiliation{Joint Research Center for Computational Biomedicine, RWTH Aachen University, Germany}
\email{zeyu.lian@rwth-aachen.de}
\author{Pieter E. Oomen}
\affiliation{Department of Chemistry and Molecular Biology, University of Gothenburg, Sweden
}
\email{pieteroomen@gmail.com}
\author{Xiulan He}
\affiliation{Department of Chemistry and Molecular Biology, University of Gothenburg, Sweden
}
\email{xiulan.he@gu.se}
\author{Soodabeh Majdi}
\affiliation{Department of Chemistry and Molecular Biology, University of Gothenburg, Sweden
}
\email{s_majdi59@yahoo.com}

\author{Andreas Schuppert}
\affiliation{Joint Research Center for Computational Biomedicine, RWTH Aachen University, Germany}
\email{schuppert@aices.rwth-aachen.de}

\author{Andrew Ewing}
\affiliation{Department of Chemistry and Molecular Biology, University of Gothenburg, Sweden
}
\email{andrew.ewing@gu.se}
\title[Classification]
  {Tree-Based Learning on Amperometric Time Series Data Demonstrates High Accuracy for Classification} 
\abbreviations{SCA, IVIEC, VIEC, ML, RF}

\begin{document}

\clearpage
\begin{abstract}
Elucidating exocytosis processes provides insights into cellular neurotranmission mechanisms, and may have potential in research on neurodegenerative diseases. Amperometry is an established electrochemical method for detection of neurotransmitters released from and stored inside cells. An important aspect of the amperometry method is the sub-millisecond temporal resolution of the current recordings which usually leads to several hundreds of gigabytes of high-quality data. In this study, we present a universal method for the classification with respect to diverse amperometric datasets using well-established data-driven approaches in computational science. We demonstrate a very high prediction accuracy ($\geq 95$\%). This includes an end-to-end systematic machine learning workflow for amperometric time series datasets consisting of pre-processing; feature extraction; model identification; training and testing; followed by feature importance evaluation - all implemented. We tested the method on heterogeneous amperometric time series datasets generated using different experimental approaches, chemical stimulations, electrode types, and varying recording times. We identified a certain overarching set of common features across these datasets which enables accurate predictions. Further, we showed that information relevant for the classification of amperometric traces are neither in the spiky segments alone, nor can it be retrieved from just the temporal structure of spikes. In fact, the transients between spikes and the trace baselines carry essential information for a successful classification, thereby strongly demonstrating that an effective feature representation of amperometric time series requires the full time series. To our knowledge, this is one of the first studies that propose a scheme for machine learning, and in particular, supervised learning on full amperometry time series data.
\end{abstract}

%%%%%%%%%%%%%%%%%%%%%%%%%%%%%%%%%%%%%%%%%%%%%%%%%%%%%%%%%%%%%%%%%%%%%
%% Start the main part of the manuscript here.
%%%%%%%%%%%%%%%%%%%%%%%%%%%%%%%%%%%%%%%%%%%%%%%%%%%%%%%%%%%%%%%%%%%%%
\clearpage
% no section title for introduction
% \section{Introduction}
% \label{sec:intro}
%%%%%%%%%%%%%%%%%%%%%%%%%%%%%%%%%%%%%%%%%%%%%%%%%%%%%%%%%%%%%%%%%%%%%%%%%%%%%%%%%%%%%%%%%%%%%%%%%%%%
% -  Group: Introduction sounds good, however, add amperometric trace schematic in a different way @done
% -  Short introduction on amperometry methods and exocytosis, highlighting the state-of-the-art papers @done
% -  Motivate the IgorPro method to analyze such data- pros and cons @done
% -  Give an overview of ML methods, recent advancements and its wide applicability, again highlighting recent interdisciplinary papers @done
% -  Few lines on literature relating ML and amperometry data, if any @done
% -  Give an overview of the supervised learning method addressed in the paper and use cases @done
% -  Some parts of the text could be borrowed from pages 1-2 of main.article.pdf @done
% -  Expected number of words: < 500 
%%%%%%%%%%%%%%%%%%%%%%%%%%%%%%%%%%%%%%%%%%%%%%%%%%%%%%%%%%%%%%%%%%%%%%%%%%%%%%%%%%%%%%%%%%%%%%%%%%%%

Neurons or nerve cells are basic units of the brain and nervous system that enable several sensory and motor functions of the body. Neurons communicate with each other through electrical signals caused by the release of neurotransmitters between neurons through exocytosis. \cite{kelly1993}. Neurotransmitters are molecules that act as chemical messengers of the body and may be excitatory or inhibitory in nature. Most neurotransmitters are small amine molecules, amino acids, or neuropeptides. Some of the well-known monoamine neurotransmitters include dopamine, norepinephrine and serotonin. Imbalance in the neurotransmitter levels may cause neurological effects leading to diseases \cite{kompoliti2010, grouleff2015}. 

Understanding the exocytosis process is vital for understanding signal transmission between neurons. Several methods have been established to quantitatively monitor neurotransmitter release, including fluorescence, mass spectrometry and amperometry methods.   
Amperometry is an established electrochemical method for the detection of
neurotransmitters released from and stored inside different cell types and have been widely used for studying the release process given their high temporal resolution \cite{liu2021}. Common applications involve chromaffin cells \cite{Dunevall2015}, beta-pancreatic cells \cite{Hatamie2020}, neuron \cite{Larsson2020} and pheochromocytoma (PC12) cells \cite{Gu2021}. The study of neurotransmitter release, or exocytosis, provides insights in underlying mechanisms of neurotransmission, as well as the pathophysiology of neurodegenerative diseases. 

The exocytosis process begins with the
the recruitment of vesicles, which carry the neurotransmitters, to the plasma membrane. Here, the vesicles partially fuse with the cell membrane and a fusion pore is created. Through this pore, neurotransmitters are then released into the extracellular environment. The released neurotransmitters may bind to receptors on target cells that initiate several intracellular signaling pathways \cite{Evanko2005, Li2016, Luzi2019}. Amperometry using microelectrodes is especially suitable for studying exocytosis, as it allows for quantification of neurotransmitters at a sub-millisecond time scale with great sensitivity. This allows monitoring of monoamine release kinetics at a single vesicular level \cite{Mosharov2005, Segura2000}. 

\begin{figure}[!htbp]
(A)\includegraphics[scale=0.5]{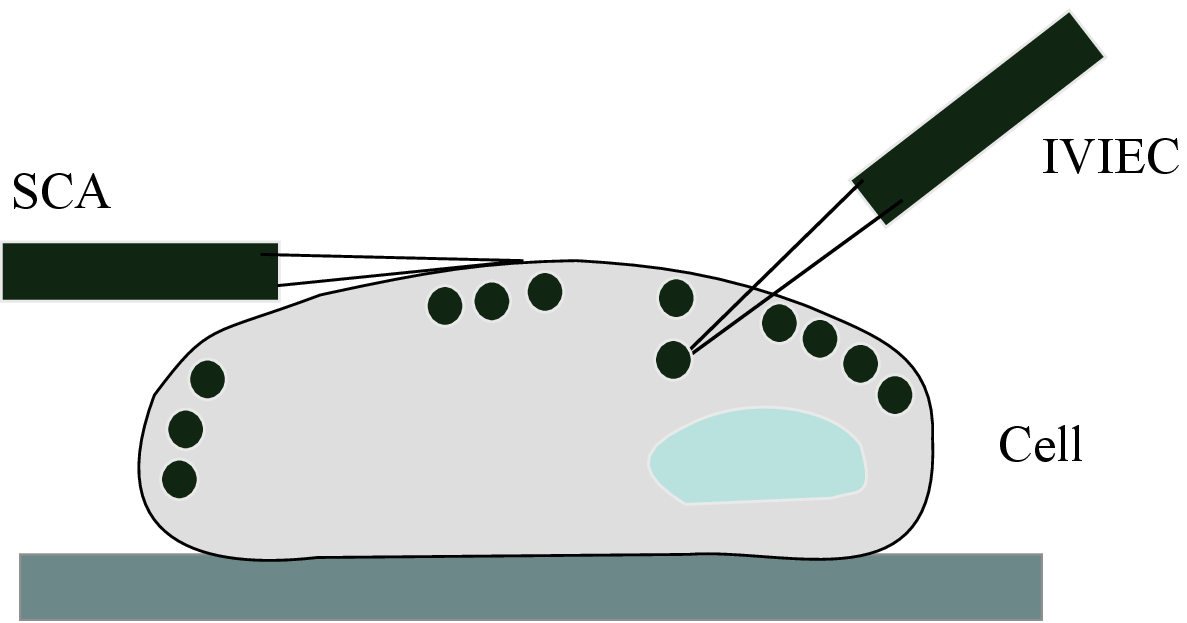}
(B)\includegraphics[scale=0.7]{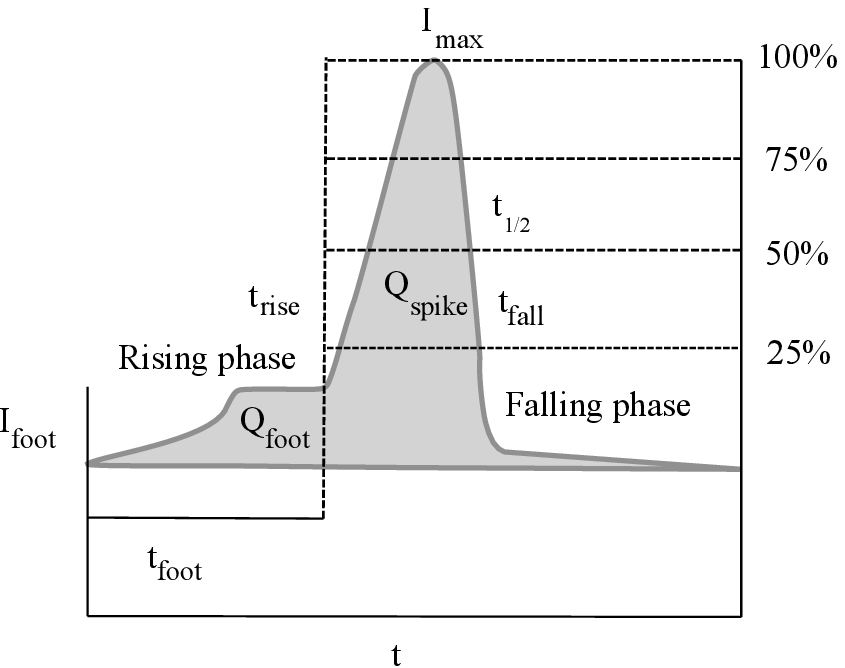} \hspace{1cm}
\caption{Illustration of methods used to measure exocytosis and the resultant amperometric spike: (A) The Single Cell Amperometry (SCA) method measures the release of neurotransmitters from the vesicles in a cell while the Intracellular Vesicle Impact Electrochemical Cytometry (IVIEC) measures the content of vesicles in the cell; another method called Vesicular Impact Electrochemical Cytometry (VIEC) measures vesicular content in a suspension of isolated vesicles (not shown in the above illustration). (B) Spike parameters: an amperometric spike is characterized by the Pre-Spike Foot (PSF), rising phase, spike, and falling phase. Here $I_\mathrm{max}$ is the peak current (w.r.t. the baseline current level), $t_\mathrm{rise}$ is the rise time (from $25$\% to $75$\% of $I_\mathrm{max}$), $t_\frac{1}{2}$ is the half peak width where $I = 25$\% $I_\mathrm{max}$, $I_\mathrm{foot}$ is the PSF current, $t_\mathrm{foot}$ is the PSF duration, $Q_\mathrm{spike}$ is the charge of the spike and $Q_\mathrm{foot}$ is the charge of the PSF.}
% \caption{Illustration of methods used to measure exocytosis and the resultant amperometric spike: (A) Single Cell Amperometry (SCA) method measures the release of neurotransmitters from a cell while the Intracellular vesicle impact electrochemical cytometry (IVIEC) measures the content of neurotransmitters in the cell; another method called Vesicular Impact Electrochemical Cytometry (VIEC) measures vesicular content in a suspension of isolated vesicles (not shown in the above illustration). (B) Spike parameters: an amperometric spike is characterized by the Pre-Spike Foot (PSF), rising phase, spike, and falling phase. Here $I_\mathrm{max}$ is the peak current (w.r.t. the baseline current level), $t_\mathrm{rise}$ is the rise time (from $25\%$ to $75\%$ of $I_\mathrm{max}$), $t_\frac{1}{2}$ is the half peak width where $I = 25\%I_\mathrm{max}$, $I_\mathrm{foot}$ is the PSF current, $t_\mathrm{foot}$ is the PSF duration, $Q_\mathrm{spike}$ is the charge of the spike and $Q_\mathrm{foot}$ is the charge of the PSF.}
\label{fig:amperometry}
\end{figure}

In Single Cell Amperometry (SCA), a microelectrode (usually carbon) is kept at a positive potential and held against the plasma membrane of a single cell. After chemical stimulation, monoamine neurotransmitters that were stored in intracellular vesicles are released from the cell. These molecules are then oxidized at the electrode surface, generating a \enquote{spike} in the current recording. Amperometric methods may monitor the quantity of neurotransmitters stored in the cell using the Intracellular Vesicle Impact Electrochemical Cytometry or IVIEC method, as well as those released from the cell using SCA (\fig\ \ref{fig:amperometry}(A)). These techniques work via the same principle: \ie\ the current that is measured over time is directly proportional to the number of molecules that is oxidized at the electrode surface. This means a comparison can be made between the average number of molecules stored in vesicles and the average number of molecules that is released by the cell. 

Because of the elaborate processes involved with exocytosis, which include docking, fusing, opening, and closing of the pores, the shapes of the obtained amperometric spikes usually differ from each other. For example, during a full fusion, the vesicle completely collapses into the plasma membrane and releases the total amount of its stored monoamines. The resulting amperometric spike is characterized by a steep rising and slow decay. In contrast, a rapidly opening and closing fusion pore would result in a sharp peak with a short decay. In addition, spikes may be simple (i.e. single peak) or complex (i.e. fluctuating peaks). The spike shapes hold crucial information on the release properties of the cell under study, in the presence or absence of external stimulants. 

For a long time, the exocytosis process was considered to be “all-or-none”. However, several recent studies have shown that it is usually a partial release process, and that cells can modulate chemical signaling by adjusting the duration and size of the fusion pore opening. It has also been shown recently that such events may be simple or complex and that, in several cell types, such partial release of neurotransmitters is followed by recycling of the vesicles for subsequent exocytosis processes \cite{Mosharov2005, Larsson2019, Hatamie, Larsson2020}.

An important aspect of the amperometry method is the temporal resolution of the so-obtained current recordings. A single recording may last up to many seconds or up to minutes with measurements taken at sub-millisecond intervals. This helps in understanding the kinetics of monoamine release (\eg\ fusion pore properties) \cite{Larsson2020, ye2018using, Majdi2017}. The amperometric experiments are also characterized by high sensitivity, which allows us to quantify released neurotransmitters and study intracellular transmitter homeostasis. Amperometric spikes are caused by the transfer of electrons after monoamine oxidize and therefore, each spike may usually correspond to an exocytotic event. 

Owing to the high resolution of recording, amperometry experimental recordings are data-rich but may produce up to several hundreds of gigabytes of data that require novel computational analysis methods for analysis. Typically a dataset is composed of several classes or categories (for example, stimulant type, cell type or type of experiment). Each class may have several measurements ranging from a few seconds to several minutes. Each measurement is a current recording (usually in pA) with respect to time. 
These measurements usually do not exhibit any periodicity and are spiky in nature. 

Given that a spike in the measurement is usually correlated with release of neurotransmitters, the state-of-the-art methods used to analyze amperometry datasets focus largely on the spike properties. These spike properties or features may include, for example, the area under the spike (related to the charge and therefore the number of molecules of neurotransmitters released during an exocytosis event and oxidizing at the electrode surface), spike frequency (related to release probability), rise time (\eg\ $25$\% - $75$\%), peak time and other similar parameters as shown in \fig\ \ref{fig:amperometry}(B).

% Due to this, the state-of-the-art methods used to analyze amperometry current data resort to investigate various properties of the spike including, but not limited to, area under the spike (related to the charge of neurotransmitters released during an exocytosis event), spike frequency (related to release probability), rise time (\eg\ $25\% - 75\%$), peak time and other similar parameters as shown in \fig\ \ref{fig:amperometry}(B). It has been shown that comparing pooled samples of all spike characteristics may lead to inconsistent results and the necessity of independent random sampling from each group of traces has been motivated \cite{Colliver2000}.

Specifically, one software program that has been widely used to study the data generated during amperometry experiments is \textit{QuantaAnalysis}(implemented in IgorPro). \textit{QuantaAnalysis} provides a framework for importing and filtering of signal; spike detection; characterization of spikes and subsequent statistical analysis wherein some steps of the analysis are highly user-dependent. Several exciting findings regarding exocytosis mechanisms for different cells under the effect of various stimulants have come out of extracting spike features from amperometric traces using IgorPro \textit{QuantaAnalysis} \cite{Mosharov2005}.

In addition, though several methods to mathematical modeling of the fusion pore expansion in the exocytosis process have been attempted \cite{Oleinick2017}, it is not complete, since we do not have full knowledge of the biochemical aspects of such processes. Such gaps in our existing knowledge could be filled efficiently by an intelligent design of experiments and subsequent empirical analysis (which may cost several years to decades) or by meticulous analysis and interpretation of already existing data (which costs a few months to a year).

Though such spike-based quantitative methods can without doubt offer insights into the exocytosis process, it is unclear whether all information related to an exocytosis process exists exclusively in their constituent spikes. 
It may well be true that information in the data exists in the baseline, spikes as well as the inter-spike dynamics. An ideal method would therefore exploit the richness of amperometry data and our knowledge of their categories or labels. In addition, such an ideal method should also have better scope for systematic standardization of the amperometry data analysis process with minimal requirement for manual intervention, thereby ensuring reproducibility.

In this regard, several data-driven methods have risen in the past couple of decades in adjacent areas of study that have generated knowledge with reasonable success. For example, in the areas of chemometrics \cite{Baronas2004, Cui2020, Namuduri2020, Gonzalez-Navarro2016, Szczesny2021, Vakilian2018, Wu2019a} (such as deep learning on electrochemical sensor data; spiking neural networks for waveform analysis in amperometry), neuroscience \cite{Glaser2020, Moore2019,  Jun2017, Karacali2010, Mulansky2016, Sarica2017, Czarnecki2015} (such as prediction of Alzheimer's disease from neuroimaging data) and in public health \cite{pmlr-v68-fiterau17a, Higgins2020, Lin2016} (such as learning longitudinal health indicators from electronic health records) among many others. Also, in the general area of electrochemical sciences, machine learning methods have begun to be widely used to analyze several types of data (including time series and imaging)\cite{SundaraRaj1999, Ma2015, Cui2020, Gu2020}. 

A recent review states that machine learning will revolutionize the electrochemical study of materials and motivates this by illustrating how switching from a trial-and-error approach of identifying useful materials for a cleaner sustainable future can be complemented by a machine learning approach that exploits the knowledge we have on the electrochemistry of existing materials\cite{Mistry2021}. Machine learning methods have been
shown to provide great predictive performance and can provide insights into novel features
that enable classification. 

In this work, we propose a machine learning method (or classifier) that achieves high accuracy for amperometric time series data. We present an end-to-end systematic workflow that includes pre-processing; feature extraction; model identification; training and testing; followed by feature importance evaluation that is implemented as a Python package. If present, dubious time series are removed during the pre-processing stage, which is then followed by chunking of the time series for straightforward feature extraction. 

The framework is based on one of the most frequently used methods in machine learning where the training samples are accompanied by labels or classes which are the target variables. We used \tsf\ for time series feature engineering. This was largely motivated by recent work on similar \enquote{spiking} data recorded from the neurons. Here, the authors use single-cell spike trains for cell and network state classification demonstrating an accuracy of up to $70\%$ using the \tsf\ features\cite{Lazarevich2018}. 

The \tsf\ Python package automatically calculates a large number of time series features from each signal and is easy to use. However, one can expect that a single model may not work on different kinds of data. The typical process to get around this aspect is to apply several methods and check which is expected to work better. We have circumvented this step by including a pre-existing Python package, the \laz, in our pipeline, which can build a lot of basic, but popular models and rank them based on expected accuracy. For the datasets that we analyzed, we identified the decision tree model or their variants to be the best classifiers with excellent performance on the metrics used.

In particular, ensemble learning algorithms such as Random Forest, are well-suited for large datasets such as that of amperometry data and can easily adapt to non-linearities found in the data\cite{Breiman2001, Breiman1996}. Using the trained classifier, new unlabelled data is passed to predict the label. Based on this, accuracy of classification is estimated and features that enabled a good classifier are used for biological interpretation.

This pipeline can perform several classification tasks. For example, it can accurately classify ion stimulations based on amperometric traces obtained from chromaffin cells. To demonstrate the universality of the proposed machine learning workflow, we use four candidate datasets acquired under different experimental conditions. The first candidate dataset (henceforth referred to as the \enquote{\hsd}), was obtained by performing SCA and IVIEC on bovine chromaffin cells that had been stimulated with different organic ions from the Hofmeister series. It renders itself directly applicable for the analysis using the proposed workflow, owing to its large size. It has been shown that there exists a link between the Hofmeister ordering of specific inorganic ions to the exocytosis process occurring in the chromaffin cells, emphasizing the role of ion specificity in the release process\cite{He2020a}. 

The second dataset (henceforth referred to as the \enquote{cells dataset}), was obtained through SCA experiments on chromaffin cells, PC12 cells\cite{Li2016}, pancreatic beta-cells, and postsynaptic terminals of muscle fibers in \textit{Drosophila melanogaster} larvae. Chromaffin cells from bovine adrenal glands and rat PC12 cells have been used extensively to study the exocytosis process of monoamine neurotransmitters\cite{Colliver2000}. Exocytosis in pancreatic beta cells has been studied as well to understand insulin secretion. \textit{Drosophila melanogaster} or fruit fly larvae have been used as a model organism to study the role of chemical neurotransmission in human neurodegenerative diseases\cite{Majdi2015}. 

The third candidate dataset (henceforth referred to as the \enquote{electrodes dataset}), was obtained through VIEC experiments using different microelectrode materials (carbon, platinum, and gold). The fourth candidate dataset (henceforth referred to as the \enquote{VIEC dataset}) was generated through VIEC experiments via the \enquote{adding} and \enquote{dipping} methods. In the dipping method, the electrode is placed in isolated vesicle suspension for roughly $20$ minutes and on ice. The electrode is then transferred to a $37$ degree Celsius isotonic buffer solution, after which a positive potential is applied to the electrode to electroporate the vesicles and record the spikes following neurotransmitter oxidation at the electrode surface. Using the adding method data is collected by placing the electrode direclty in the isolated vesicle suspension at $37$ degree Celsius and applying the positive potential.

The choice of heterogeneous datasets has been deliberate, so as to illustrate the versatility of the proposed pipeline in handling different experimental methods, cell sources, and stimulations involved in the study of exocytosis.

\section*{Results and Discussion}
\label{sec:res}

%%%%%%%%%%%%%%%%%%%%%%%%%%%%%%%%%%%%%%%%%%%%%%%%%%%%%%%%%%%%%%%%%%%%%
% -  Group: IVIEC/VIEC method - good to add a schematic showing them @done
% -  Group: Good to make piechart as histogram with readable names instead of variable strings @done
% -  Motivate candidate datasets- Hofmeister, electrodes, Cells, IVIEC/VIEC @done
% -  Emphasize the diversity of methods and categories @done
% -  Show accuracy achieved in these methods (similar to the table in the presentation or Figure 7) @done
% -  Show histogram of top features for (a) one candidate dataset (b) across different datasets (c) across different models @done
% -  Motivate on a few features, explain the importance value distribution and future scopes for this method; ISI stuff @done
% -  Optional: images of raw data and some words on the preliminary IgorPro results  @done
% -  Expected number of figures: 3-4
% -  Expected number of words: < 500
%%%%%%%%%%%%%%%%%%%%%%%%%%%%%%%%%%%%%%%%%%%%%%%%%%%%%%%%%%%%%%%%%%%%%
\subsection*{Classifier}
\label{subsec:classifier}

As motivated in the introduction, a machine learning algorithm or a classifier in machine learning is an algorithm that is capable of categorizing data into a class or label. A classifier is then the algorithm itself, that is used to classify data. Several types of classifiers exist, however, they are usually broadly classified as supervised and unsupervised machine learning classifiers. Unsupervised machine learning classifiers are capable of identified patterns or structures in an unlabelled dataset. On the other hand, supervised machine learning classifiers learn to classify data based on predetermined categories from a labelled dataset. Given that the amperometry datasets are data-rich and labelled, we opt for supervised machine learning classifiers, such as decision tree models, in our analysis.
%%%%%%%%%%%%%%%%%%%%%%%%%%%%%%%%%%%%%%%%%%%%%%%%%%%%%%%%%%%%%%%%%%%%%
\subsection*{Hyperparameter Tuning}
\label{subsec:hyperparams}

Machine learning algorithms require several settings of several high-level parameters for training a learning algorithm. These parameters are usually left for the user to pick and are generally referred to as hyperparameters. In the case of tree-based classifiers, such hyperparameters could be, for example, the number of decision trees in the forest, depth of indvidual decision trees, number of splits at every node of a tree, criteria for splitting, among many other hyperparameters. It is usually hard to estimate exact hyperparameter settings for training and requires initial trial-and-error approaches from the user, followed by use of simple search algorithms (such as grid search) for hyperparameter tuning or intelligent algorithms (such as Bayesian search) to identify the best set of hyperparameters. For a more detailed description, please see the methods in \sect\ \ref{sec:methods}.
%%%%%%%%%%%%%%%%%%%%%%%%%%%%%%%%%%%%%%%%%%%%%%%%%%%%%%%%%%%%%%%%%%%%%
\subsection*{Classification using \tsf}
\label{subsec:tsf}

As introduced in the previous section, we used four candidate datasets for the analysis of the classification workflow. These candidate datasets include \begin{enumerate*} \item the \hsd\ generated through SCA experiments wherein the classes are the Hofmeister ions including Cl$^-$, Br$^-$, NO$_{3}^{-}$, ClO$_{4}^{-}$ and SCN$^{-}$; \item the cells dataset also generated through SCA experiments wherein the classes are the beta-cells, chromaffin cells, PC12 cells and neuromuscular endplates in \textit{D. melanogaster} larvae; \item the electrodes dataset generated through VIEC experiements using different microelectrode materials (gold, platinum, carbon); \item the VIEC dataset whose classes include the methods used to monitor the vesicular content, \ie\ the \enquote{adding} and \enquote{dipping} methods \end{enumerate*}. Each of these datasets went through pre-processing; chunking; feature extraction; best classifier identification; followed by training and prediction. The detailed workflow for each of these steps is described in the methods section.

%%%%%%%%%%%%%%%%%%%%%%%%%%%%%%%%%%%%%%%%%%%%%%%%%%%%%%%%%%%%%%%%%%%%%
% These candidate datasets include \begin{enumerate*} \item \hsd\ dataset generated through SCA experiments wherein the classes are the Hofmeister ions including Cl$^-$, Br$^-$, NO$_{3}^{-}$, ClO$_{4}^{-}$ and SCN$-$ \item Cells dataset also generated through SCA experiments wherein the classes are the Beta cells, Chromaffin cells, PC12 cells and cells from the fruit fly (\textit{D. Melanogaster}) \item Electrodes dataset generated through SCA experiments whose classes are the types of electrodes used in the experiment including carbon, platinum, and gold \item VIEC dataset whose classes include the methods used to monitor the vesicular content, i.e. adding and dipping method \end{enumerate*}. 
%%%%%%%%%%%%%%%%%%%%%%%%%%%%%%%%%%%%%%%%%%%%%%%%%%%%%%%%%%%%%%%%%%%%%
\subsection*{Key Observations}
\label{subsec:keyobs}

The key takeaways from the proposed pipeline are the following. Firstly, it is important to note that the
features used in the classification workflow are critical. Specifically, using spike-centric features exclusively will not lead to a good classifier. This was validated through our first analysis of the pipeline using \textit{QuantaAnalysis} features that were extracted from spikes exclusively. These included about ten spike features including the time, current, and PSF parameters. An accuracy of only $\approx 26\%$ was achieved with a very high standard deviation which indicates that spike features alone are insufficient to clearly distinguish classes from one another (please refer to methods section for detailed information).

Secondly, using a dedicated time series feature extraction library such as the \tsf\ on full-time series leads to a good classifier for all the candidate datasets. Concretely, for all candidate datasets, tree-based classifiers were found to offer the best accuracy. Among these, several tree-based classifiers, including Random Forest, Extra Trees, and XGBoost, emerged as the best classifiers which were used for prediction. For all combinations of datasets and tree-based classifiers (\ie\ four datasets and three models), we could get an accuracy of $\geq 95\%$. This shows that the combination of selected features contributes to a high certainty of distinction from one class to another. 

Thirdly, such high accuracy is observed despite the diversity of the analyzed amperometric datasets. Specifically, these datasets have been generated through different experiments, chemical stimulations, electrodes, and varying recording times. However, almost invariably, tree-based methods turn out to be the best classifiers for these datasets with very high accuracy. This shows that the combination of selected features contributes to a high certainty of distinction from one class to another.
%%%%%%%%%%%%%%%%%%%%%%%%%%%%%%%%%%%%%%%%%%%%%%%%%%%%%%%%%%%%%%%%%%%%%
\subsection*{Importance Values}
\label{subsec:importance}

In a supervised machine learning problem such as that of the classification of amperometry datasets, it can be expected that not all featurues extracted by \tsf\ played an equal role. In other words, some features strongly enabled the classifier to predict on a test dataset more than the others, thereby giving insights into the model and improvements. In addition, these top features are more likely the ones that require our attention for biological interpretation. In this regard, feature importance methods refers to techniques for assigning scores to input features of a model that indicate the relative importance of each feature when making a prediction. With respect to tree-based classifiers, several feature importance methods such as Gini importance and permutation-based feature importance exist among many others, and we point the curious reader to the related references \cite{Breiman2001, Breiman1996}.
%%%%%%%%%%%%%%%%%%%%%%%%%%%%%%%%%%%%%%%%%%%%%%%%%%%%%%%%%%%%%%%%%%%%%
\subsection*{Top features}
\label{subsec:top}

\begin{figure}[!htbp]
\centering
\textbf{(A)}\includegraphics[scale=0.33]{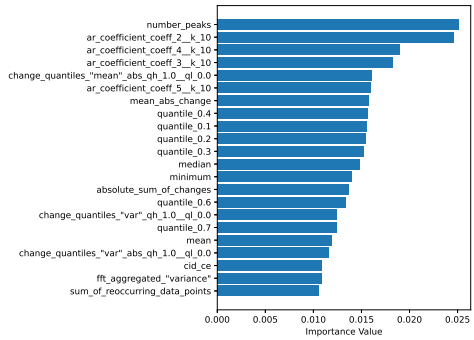} \hspace{0.5cm}
\textbf{(B)}\includegraphics[scale=0.33]{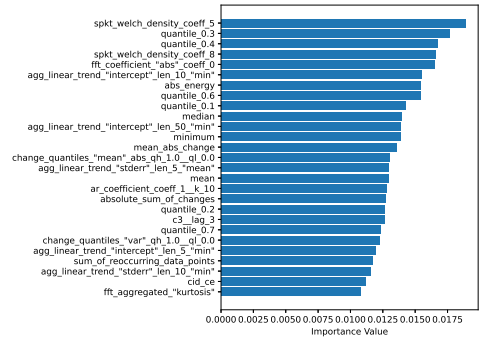}\\
\textbf{(C)}\includegraphics[scale=0.33]{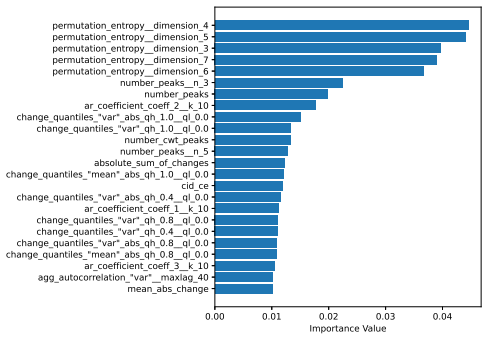} \hspace{0.5cm}
\textbf{(D)}\includegraphics[scale=0.33]{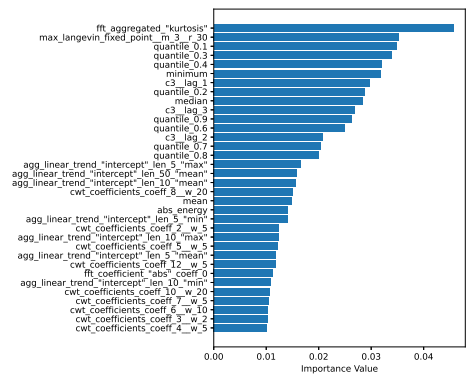}\\
\caption{Histogram of the top features (with importance value over $1\%$) extracted from (A) \hsd. (B) Electrodes dataset. (C) Cells dataset. (D) VIEC dataset. The labels of the top features are explained in detail in the supplementary section. The names of top features have been left identical to that of \textit{tsfresh} documentation for consistency and easy comparison.}
\label{fig:hist}
\end{figure}

There was no single \tsf\ feature with a remarkably high importance value that directly assisted the classifier in prediction. Instead, the importance values were thinly distributed among the extracted features, where most importance values are $\leq 0.01$. Features having importance values slightly greater than $0.01$ can be captured by applying the one percent threshold as can be seen in \fig\ \ref{fig:hist}. We encourage the interested reader to refer to the supplementary section for more information on these top features that emerged at the top one percent for the individual datasets along with a few other auxiliary features. 

Some of these top features (\eg\ the number of peaks) are simple statistical metrics, while others (\eg\ permutation entropy) measure complexity in the time series. Similarly, some of these features are time-domain metrics while others are metrics in the frequency domain. Furthermore, the datasets do not share their top most features. For example, for the \hsd, the top most feature is a time-domain feature that calculates the number of peaks (i.e. local maxima in the mathematical sense) in the input time series. Peaks indicate critical events such as an increase in neurotransmitters release during the exocytosis process. Though small peaks are not considered as an emergent property of an ongoing process, and perhaps, usually ignored, this feature seems to be a strong indicator for differentiating class labels in the \hsd. While it is a trivial task to identify small peaks visually, \tsf\ uses a formal definition for identifying such small peaks. A data-point is identified as a peak when it is a local maximum in a given local time window parameterized by its half-window size, however, it is not large nor globally maximum in the entire time series. Further, a peak is usually isolated in the sense that not too many points in that time window have similar values.
% auto-regression coefficient is a close first

However, the top most feature for the electrodes dataset is a feature in the frequency domain \ie\ the spike welch density coefficient. It estimates the cross-spectral density of an input time series at different frequencies. This feature calculates the power spectral density of a given time series by dividing it into multiple overlapping segments, identifying dominant frequencies for each segment, and then averaging over them. % quantile is a close first 
Permutation entropy surfaces as the top few most important feature, which is again a complexity measure. It measures the complexity of a dynamic system by capturing the order relations between values of a time series and extracting a probability distribution of the ordinal pattern. For the VIEC dataset, it is a frequency-domain measure that emerges as a top feature \ie\ the kurtosis of the absolute Fourier Transform.

Taking a bird's eye view of all top features in all the datasets, we can make a straightforward inference that several common simple time-series statistics play a key role in the classification task as well. Measures such as mean, median, quantile, and a minimum of the time series emerge as critical time-domain features across all datasets. Few other high-level features such as change\_quantiles (a feature that is calculated by fixing a window of the time series values and then calculating a set of consecutive change values in the series and then applying an aggregation function); fft\_coefficient (Fast Fourier Transform Coefficients) and agg\_linear\_trend: features from the linear least-squares regression for the values of the time series that are aggregated over chunks of a certain size; all seem common recurring features of the feature matrix.

At an even higher level, we identified some common top features across several parts of this analysis presented here, including across the four datasets, and across the three methods used for each of the datasets summarized in \tab\ \ref{tab:top}. Across different amperometric time series datasets which we analyzed, two types of features were common among the top $0.3$\% of the features. The first is the c3 statistics, which is a measure of non-linearity in time series, namely the third-order autocovariance, and the second is a simple statistical measure: quantile.

% At an even higher level, we could identify some common top features across several parts of this analysis presented here including \begin{enumerate*} \item across the four datasets \item across the three methods used for each of the dataset\end{enumerate*} summarized in \tab\ \ref{tab:top} to which we will pay most attention here in the remaining part of this section. Across different amperometric time series datasets that we analyzed, two types of features were common among the top $0.3\%$ of the features. The first is the c3 statistics which are a measure of non-linearity in time series, namely the third-order autocovariance and the second is a simple statistical measure, quantile. 

For the three classifiers used for the \hsd, apart from the common time-domain features (such as quantiles, median, and the number of peaks), auto-regression coefficients (which allows the regression on the current value at succeeding time steps using history values) and mean absolute change (total variation) emerged as common dominant features. Mean absolute change is also a time-domain metric, an average over first differences, and is calculated from subsequent values in a time series. The coefficients of an Auto-regressive model are also calculated in the time domain and are used to describe time-varying processes through a stochastic differential equation. For the other datasets, similar simple features such as minimum and median emerged along with complex time-domain features such as permutation entropy and frequency-domain features such as Continuous Wavelet Transform coefficients and FFT aggregated kurtosis. 

It is important to note that, in a parallel analysis, we used this supervised learning pipeline for learning on the Inter-Spike Interval (ISI) data wherein we extracted the spike times based on a pre-defined threshold from these candidate datasets and classified the data based on the series of interspike interval lengths. However, the classification accuracy achieved was inadequate ($22.38\%$ with $7.31\%$ standard deviation) to make a good classifier. This reinforced the claim that the information in an amperometric time series dataset exists not exclusively in spike shape nor their triggering pattern but in the full-time series. 

\clearpage
\begin{sidewaystable}
\begin{tabular}{ |p{4cm}|p{6cm}|p{4cm}|p{4cm}|p{4cm}| } 
\hline
\textbf{Across datasets (for Random Forest Classifier) threshold $= 0.3\%$} &
\textbf{Across methods (Hofmeister) threshold $= 1\%$} & \textbf{Across methods (electrodes) threshold $= 1\%$} & \textbf{Across methods (cells) threshold $= 1\%$} & \textbf{Across methods (IVIEC/VIEC) threshold $= 1\%$}\\   \hline  \hline
c3\_lag\_3 &  number\_\newline peaks\_n\_1& minimum & permutation\_\newline entropy\_dimension\_6\_\newline tau\_1& fft\_aggregated\_\newline aggtype\_kurtosis\\  \hline
quantile\_q\_0.7& quantile\_q\_0.7& permutation\_\newline entropy\_dimension\_4\newline \_tau\_1& quantile\_q\_0.4 & \\  \hline
c3\_lag\_1& quantile\_q\_0.4 & permutation\_\newline entropy\_dimension\_7\newline\_tau\_1& &\\
c3\_lag\_2& quantile\_q\_0.6 & & &\\  \hline
quantile\_q\_0.1& mean\_abs\_change& & &\\  \hline
& median &  & &\\  \hline
& ar\_coefficient\_coeff\_4\_k\_10 & & &\\  \hline
& quantile\_q\_0.8 & & &\\  \hline
\end{tabular}
\caption{List of top features that are common (A) across different datasets for analysis using the Random Forest method (Importance threshold $=0.3\%$), (B) across different decision tree methods for the Hofmeister Series dataset (Importance threshold $=1\%$) (C) across different decision tree methods for the cells dataset (Importance threshold $=1\%$) and (D) across different decision tree methods for IVIEC/VIEC dataset (Importance threshold $=1\%$). The labels of the top features are explained in detail in the supplementary section. The names of top features have been left identical to that of \textit{tsfresh} documentation for consistency and easy comparison.}
\label{tab:top}
\end{sidewaystable}
\clearpage

%%%%%%%%%%%%%%%%%%%%%%%%%%%%%%%%%%%%%%%%%%%%%%%%%%%%%%%%%%%%%%%%%%%%%%%%%%%%%%%%%%%%%%%%%%%%%%%%%%%%%%%%%%%%%%%%%%%%%%%%%%%%%%%%%%%%%%%%%%%%%%%%%%%%%%%%%%%%%%%%%%%%%%%%%%%%%%%%%%%%%%%%%%%%%%%%%%%%%%%%%%%%%%%%%%%%%%%%%

%%%%%%%%%%%%%%%%%%%%%%%%%%%%%%%%%%%%%%%%%%%%%%%%%%%%%%%%%%%%%%%%%%%%%
\clearpage
\section*{Conclusions}
\label{sec:conclu}

%%%%%%%%%%%%%%%%%%%%%%%%%%%%%%%%%%%%%%%%%%%%%%%%%%%%%%%%%%%%%%%%%%%%%
% -  Can we have a separate conclusions section?
% -  Should it be after methods?
% -  Summarize paper contribution
%%%%%%%%%%%%%%%%%%%%%%%%%%%%%%%%%%%%%%%%%%%%%%%%%%%%%%%%%%%%%%%%%%%%%
We have outlined a universal method to analyze diverse amperometric datasets using well-established data-driven approaches in computational science. Using a standard machine learning pipeline, we have demonstrated that using several hundred features calculated from full-time series data can enable very high accuracy prediction and data classification tasks on amperometric time series. There is no single dominant feature that has a very high importance value, however, we have identified a certain overarching set of features that comes out of analyzing different kinds of amperometry data (obtained from different cells, electrodes, chemical injections across different experiments) with very high accuracy ($>95\%$). These features range from simple statistical features in the time domain and frequency domain to several complexity measures. The time cost for feature extraction and classification runs into a few hours depending on the volume of the data. Further, we have shown that a subset of these features emerges independently of the classifier used to learn the data. 

The proposed workflow and corresponding observations are generalizable to several datasets. The choice of the classifier was also motivated by the highly parallel structure, where each tree can be independently trained and evaluated, leading to an optimal training time due to the simplicity of the constituent decision tree classifiers. Further, the classifier relies completely on the characteristics of the datasets and makes few or no assumptions on the underlying exocytosis process. 

At this point, it is also important to note that an effective feature representation of amperometric time series requires the full-time series. It is not necessarily in the spikes alone nor in the temporal structure of spiking but also in the transients between spikes. This is a key finding, that does not align with the premise of several state-of-the-art amperometry methods that focusses largely on spike properties alone. 
Further, the proposed method requires that there is sufficient data to learn the features necessary for prediction. In other words, if the datasets have too few samples per class, in particular with short recording times then it may be challenging to achieve high accuracy. This may lead to a poor prediction accuracy with decision tree in the proposed pipeline.

The machine learning pipeline, however, is not immune to hyperparameter tuning. Hence, it may be useful to incorporate methods that can automatically do a systematic hyperparameter search for the proposed workflow. Further, Random Forest classifiers cannot express combinations of features, and hence, it is possible that there exists correlations between the identified features. However, in our preliminary analysis of the top one percent of the features that came out of the \hsd\ no correlations were found. In addition, the performance of tree-based classifiers is not affected by correlating features, if any. Nevertheless, identifying key hidden parameters that influence such correlated features is an open challenge.

%%%
% The developers of the tsfresh package did not have statistical correlation (or lack of correlation thereof) in mind when designing the features. Many of them are contributed by the community- so there is no initial thought process behind them. If correlation is a concern, the users are suggested to perform e.g. PCA to retrieve a linear independent set of features, which however does not significantly improve the performance according to the developers.

% During the feature selection, correlation does not play a huge role, as the advanced users select features one-by-one. For a more sophisticated selection process that takes correlation among features into account, we would advise using a robust ML model - such as a BDT and have a look into the feature importance it generates.

% As for tree-based classifiers used in the present paper, they are immune to the multicollinearity problem, i.e. correlating features won't impact the performance. However, readers who intend to try out other classification architectures should bear the possible feature correlations in mind.
%%%

A strong biological interpretation of the features generated from this analysis requires the development of a concrete hypothesis arising from observations of one or more of these features, followed by the design of experiments and subsequent validation. Hence, this study opens a new way for reverse feature engineering of the most relevant features that may aid in uncovering partially unknown exocytosis mechanisms. Certainly, a multi-pronged strategy of meticulous data analysis, intelligent design of experiments, and model development would accelerate the pace of research in understanding the exocytosis process.

To our knowledge, this is one of the first studies that propose a pipeline for machine learning on full amperometry time series data, in particular, using supervised learning methods. Given the volume and variety of kinds of exocytosis data produced, opens an avenue for applying machine learning methods on electrochemical data in the amperometry community. This is further facilitated by the open-source access to machine learning model implementations available online with relatively straightforward usability.
%%%%%%%%%%%%%%%%%%%%%%%%%%%%%%%%%%%%%%%%%%%%%%%%%%%%%%%%%%%%%%%%%%%%
%%%%%%%%%%%%%%%%%%%%%%%%%%%%%%%%%%%%%%%%%%%%%%%%%%%%%%%%%%%%%%%%%%%%
\clearpage
\section*{Methods}
\label{sec:methods}

%%%%%%%%%%%%%%%%%%%%%%%%%%%%%%%%%%%%%%%%%%%%%%%%%%%%%%%%%%%%%%%%%%%%
% -  Group: motivate the general method here, however go deeper in supporting information @done
% -  Perhaps add a representative schematic of the exocytosis method and/or time series feature extraction? @done
% -  Method pipeline (similar to Figure 5 in main.article.pdf and box) @done
% -  Table to expand on the different methods @done
% -  Computational cost/ chunk sizes (floats and text can be adapted from pages 5-6) and page 14 @done
% -  Expected number of floats: 5-7 (5 figures + 2 tables) 
% -  Expected number of words: < 500
%%%%%%%%%%%%%%%%%%%%%%%%%%%%%%%%%%%%%%%%%%%%%%%%%%%%%%%%%%%%%%%%%%%%

\subsection*{Classification using \textit{QuantaAnalysis}}
\label{subsec:quanta}

As motivated in the previous sections, amperometry traces are high-resolution one-dimensional time-series data that make for ideal candidates to be treated as a classification problem in the area of machine learning. The traditional analysis is usually done by using a mature spike-based amperometric data analysis software, IgorPro \textit{QuantaAnalysis}. The \textit{QuantaAnalysis} software characterizes the amperometric trace through the characteristics of the constituent spikes isolated from the traces based on a pre-defined threshold. As a first step, we extracted these spike features from the raw signal of a candidate dataset (the \hsd) and identified the best model that gives good accuracy for these features. However, for the resulting $10-$dimensional feature space, with a training feature matrix of shape $(113, 10)$, the accuracy of the best model was $\leq 30\%$ (first row of \tab\ \ref{tab:shape_acc}). This showed that, even though it is possible to retrieve Hofmeister-ordering through analysis of exclusively these spike characteristics by doing time-domain analysis\cite{He2020a}, these were insufficient to make a good classifier. This is caused by significant overlaps in the feature distributions among different classes. Hence, classifying these datasets using standard spike-based features alone is not possible.

%%%%%%%%%%%%%%%%%%%%%%%%%%%%%%%%%%%%%%%%%%%%%%%%%%%%%%%%%%%%%%%%%%%%
\subsection*{Feature Extraction Libraries}
\label{subsec:libraries}

As we would like to extract full time series features, this strongly motivates the use of time series feature extraction libraries that did not rely only on the spike characteristics of the amperometric traces. Therefore, we used \textit{\tsf}, a Python package that automatically calculates a large number of time series characteristics to do feature extraction. It is important to note that, unlike \textit{QuantaAnalysis}, \textit{\tsf} calculates features from the full-time series and not just the spikes. \textit{\tsf} calculates $\approx 794$ features including basic statistical features such as means, medians, quantiles among many others. Advanced features that are also calculated include Fast Fourier Transform (FFT) features and those extracted from more complex mathematical models, such as autoregression coefficients for each time series. Features can be filtered or extracted based on the importance values for a given classification task. Using the features, models with the best accuracy were identified using \laz, a Python package that helps evaluate the most commonly used basic ML models and understand which model works better for a given dataset without any parameter tuning. For more detailed information on these packages, please refer to the supplementary section.

%%%%%%%%%%%%%%%%%%%%%%%%%%%%%%%%%%%%%%%%%%%%%%%%%%%%%%%%%%%%%%%%%%%%
\subsection*{Overview of the Tree-Based Learning Workflow}
\label{subsec:tree}

Since \tsf\ package can extract key features for time-series classifications, the same set of features were used to evaluate many classification methods \eg\ logistic regression, Support Vector Machines (SVM), among many others. At a bird's-eye view, workflow for the analysis of the amperometric dataset involves several steps, including pre-processing, feature extraction, splitting of train and test sets followed by training and prediction as illustrated in \fig\ \ref{fig:workflow}. First, the raw data is pre-processed to remove dubious time series or unrealistic values, followed by chunking, where the full-time series is subdivided into small chunks with fixed chunk sizes due to computational reasons. This is followed by feature extraction using \tsf, wherein a feature vector with up to $794$ dimensions is generated for each chunk. Subsequently, a set of filtered features are extracted based on their importances to the classification task, which is then split into train and test sets (usually with an $80/20$ ratio). Note that the filtered features are usually much fewer than the full features set.

%%%%%%%%%%%%%%%%%%%%%%%%%%%%%%%%%%%%%%%%%%%%%%%%%%%%%%%%%%%%%%%%%%%%
\subsection*{Training}
\label{subsec:train}
Training is a critical step in machine learning where the model learns to identify similar time series. It also helps finding optimal hyperparameters of the model. The best model was identified using the \laz\ package. Here we note that for all the candidate datasets, tree-based models showed the highest accuracy. The decision tree classifier (variants shown in the box in \fig\ \ref{fig:workflow}) is trained using the training set, followed by prediction on the testing set. The classification accuracy, calculated from the number of correct predictions, defines the goodness of a classifier. A good classifier enables us to make several inferences including the most relevant features that aided in the classification problem. These relevant features may give further insights on the biological interpretation of the top features that may eventually enable a better understanding of the exocytosis process (more on this in the results section).

%%%%%%%%%%%%%%%%%%%%%%%%%%%%%%%%%%%%%%%%%%%%%%%%%%%%%%%%%%%%%%%%%%%%
\subsection*{Chunking}
\label{subsec:chunk}
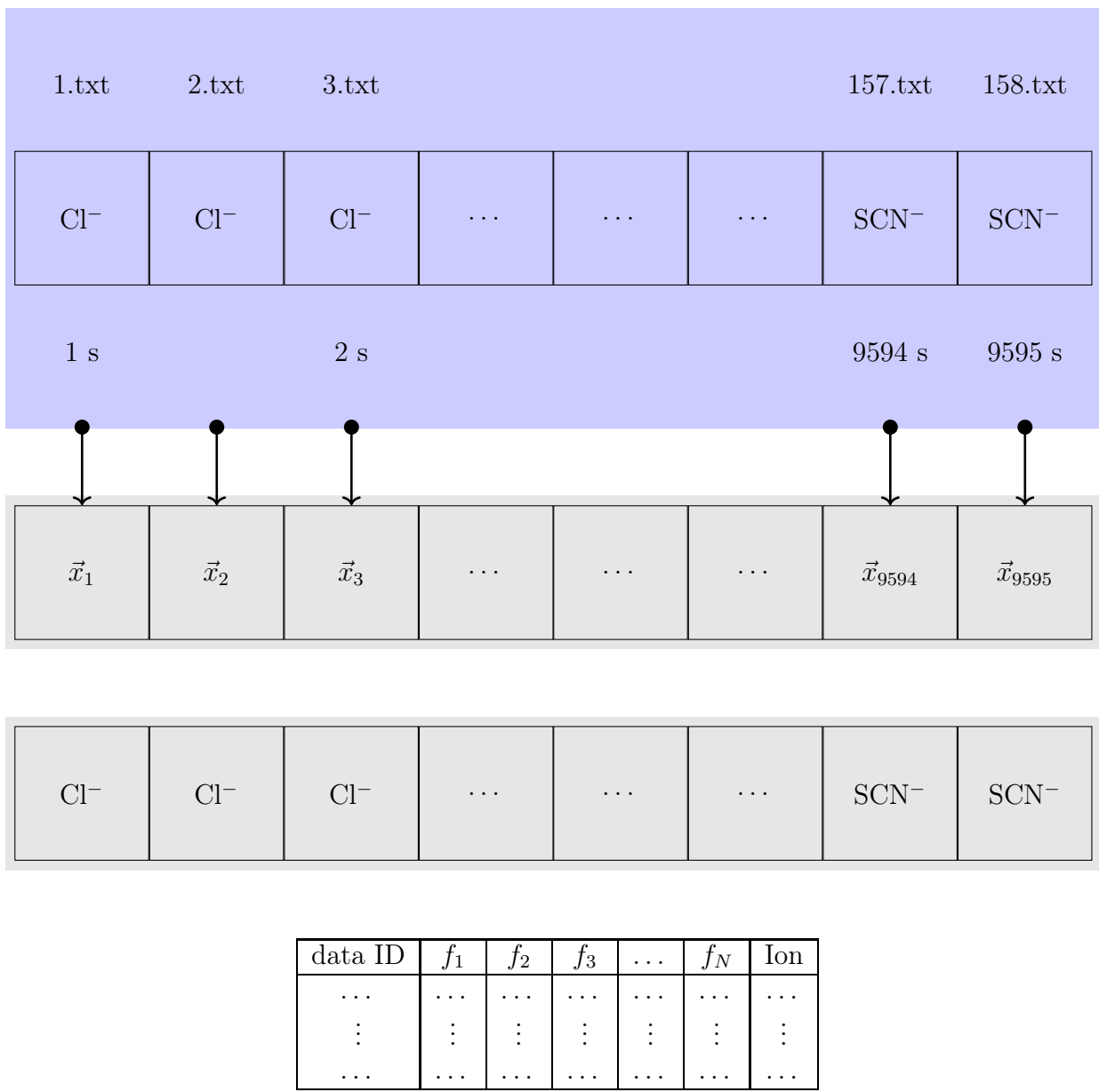
\begin{figure}[!htbp]
\centering
\begin{tikzpicture}
\matrix[matrix of nodes, fill=blue!20, nodes={draw, minimum size=20mm,anchor=center},nodes in empty cells, minimum height = 1cm,row 1/.style={nodes={draw=none}},row 3/.style={nodes={draw=none}},] at (0,0) % label=east:Data,
(mat1)
{
$1.$txt & $2.$txt & $3.$txt &&&& $157.$txt & $158.$txt\\
Cl$^-$&  Cl$^-$&  Cl$^-$& \ldots & \ldots & \ldots & SCN$^-$& SCN$^-$ \\
$1$ s &  & $2$ s &&&& $9594$ s & $9595$ s\\
};

\matrix[below = 1cm of mat1, fill=gray!20, matrix of nodes,nodes={draw, minimum size=20mm,anchor=center},nodes in empty cells,minimum height = 1cm] % label=west:$\vec{x}_i$, label=east:Feature Extraction,
(mat2)
{
$\vec{x}_1$ & $\vec{x}_2$ & $\vec{x}_3$ & \ldots & \ldots & \ldots & $\vec{x}_{9594}$ & $\vec{x}_{9595}$\\
};

\matrix[below = 1cm of mat2, fill=gray!20, matrix of nodes,nodes={draw, minimum size=20mm,anchor=center},nodes in empty cells,minimum height = 1cm] % label=west:$y_i$,
(mat3)
{
 Cl$^-$&  Cl$^-$&  Cl$^-$& \ldots & \ldots & \ldots & SCN$^-$& SCN$^-$ \\
};

\begin{scope}
\draw[*->, line width=1pt]
  (mat1-3-1.south) -- (mat2-1-1.north);
\draw[*->, line width=1pt]
  (mat1-3-2.south) -- (mat2-1-2.north);
\draw[*->, line width=1pt]
  (mat1-3-3.south) -- (mat2-1-3.north);
\draw[*->, line width=1pt]
  (mat1-3-7.south) -- (mat2-1-7.north);
\draw[*->, line width=1pt]
  (mat1-3-8.south) -- (mat2-1-8.north);
% \draw[*->, line width=1pt,shorten <=0pt,shorten >=0pt]
  % (mat2-1-4.south) -- (mat3-1-4.north);
\end{scope}
\end{tikzpicture} 
\vspace{1cm}\\
\begin{tabular}{ |c|c|c|c|c|c|c| } 
 \hline
 data ID & $f_1$ & $f_2$ & $f_3$ & \ldots & $f_N$ & Ion\\
 \hline
 \ldots & \ldots & \ldots & \ldots & \ldots & \ldots & \ldots\\
  \vdots & \vdots & \vdots & \vdots & \vdots & \vdots & \vdots\\
  \ldots & \ldots & \ldots & \ldots & \ldots & \ldots & \ldots\\
 \hline
\end{tabular}
\caption{Illustration of the time series feature extraction: Top row: Raw data of amperometric time series received as .txt files with corresponding category labels (herein shown as the Hofmeister ions as an example) are pre-processed and chunked into several samples (here shown as chunked at one-second resolutions). Middle row: Features, $\vec{x}_i$ where $\vec{x}_i \in R^{794}$ are extracted for each of these chunked time series. Last row: Corresponding labels, $\vec{y}_i$ (here shown as the Hofmeister ions as an example) are recorded. This gives the feature matrix along with the ion labels shown at the bottom. A single file may have up to several seconds of data.}
\label{fig:chunking}
\end{figure}

Amperometric traces have a very high temporal resolution with lengths of the order of $\sim 10^5-10^6$. Hence, feature extraction from such large time series, which cannot be parallelized at the data level, is not computationally feasible. However, the parallelization at the feature level is possible but may not be relevant here. Hence we resort to a method called chunking, wherein we segment large time series into segments of up to one-second size (or $10^4$ time points at $10$kHz sampling frequency) as illustrated in \fig\ \ref{fig:chunking}. Consider the case where we have been given multiple data files (here referred to as numbered text files). Each of these files may have time series that may contain several hundred thousands to millions of sampling points. Each of these text files carries a pre-defined category label (such as the ion label from \hsd\ data denoted in the figure) which may correspond to tens or hundreds of seconds. One can imagine that, for computational reasons explained above, these files can be chunked into shorter segments of data from which extraction of the feature vector, $\vec{x}_i$, becomes computationally more efficient and parallelizable (shown in the middle row of the figure). Further, we can retain the label mapping, $y_i$ (shown in the last row of the figure), which is essential for the supervised learning task. This way, a feature matrix can be constructed which consists of information of the feature vector and the corresponding label (in this case, ion) and their corresponding data ID (filename), as represented in the last block of \fig\ \ref{fig:chunking}. The feature matrix for each class is therefore of dimensions \# total number of chunks within class $\times$ number of filtered features. The advantages of chunking are two-fold: \begin{enumerate*} \item chunking significantly accelerates the feature extraction step through parallelization and avoids memory overflow, and \item chunking increases the amount of training samples belonging to a category, \ie\ the training set size. \end{enumerate*}

% \begin{enumerate*} \item Chunking significantly accelerates the feature extraction step through parallelization and avoids memory overflow \item Chunking increases the amount of training samples belonging to a category \ie\ the training set size \end{enumerate*}.

%%%%%%%%%%%%%%%%%%%%%%%%%%%%%%%%%%%%%%%%%%%%%%%%%%%%%%%%%%%%%%%%%%%%
\subsection*{Decision Tree-based Classifiers}
\label{subsec:decision_tree}

\begin{figure}[!htbp]
\begin{forest}
for tree={
  grow=east,
  draw=blue!50,
  fill= gray!20,
  circle,
  minimum size=2em,
  line width=2pt,
  parent anchor=east,
  child anchor=west,
  edge={draw=black},
  edge label={\Huge\color{black}},
  edge path={
    \noexpand\path[\forestoption{edge}]
      (!u.parent anchor) -- ([xshift=-1.6cm].child anchor) --    
      (.child anchor)\forestoption{edge label};
  },
  l sep=2cm,
}, 
tikz+={
\tikzstyle{arrow}=[thick,->,>=stealth]
\tikzstyle{sblock} = [rectangle, draw, 
text width=3cm, text centered, rounded corners, minimum height=1cm]
\draw [black, thick] (10.8,-7) to [square right brace] (10.8,7);
\draw [arrow] (11.3,0) --  (12.5,0);
\node[draw,circle,minimum size=1cm,inner sep=0pt] at (13,0) {$\Sigma$};
\draw [arrow] (13.5,0) --  (14.7,0);
\node[draw, align=left] at (15.75,0) {Final\\ Prediction};
\node[align=left] at (15.75,-1) {\scriptsize Class C};
\node[align=left] at (7,8) {Prediction};
\node[align=left] at (13,8) {Majority\\ Voting};
\node[align=left] at (0.1,8) {Data};
\draw[black, dashed] (2,7) rectangle (9.5,2.5);
  },
[D,rectangle, s sep=10pt
  [,edge label={node[Below]{$T_N$}}
    [,edge label={node[Below]{}}
    [,edge label={node[Below]{}}
    ]
    [,edge label={node[Above]{}}
    ]
    ]
    [,edge label={node[Above]{}}
    [,edge label={node[Below]{}}
    ]
    [,label=Class C,edge label={node[Above]{}}
    ]
    ]
  ]
  [,label = \vspace{-2cm}\longvdots{3em},edge label={node[Above]{$T_2$}}
    [,edge label={node[Below]{}}
    [,edge label={node[Below]{}}
    ]
    [,edge label={node[Above]{}}
    ]
    ]
    [,edge label={node[Above]{}}
    [,edge label={node[Below]{}}
    ]
    [,label=Class A,edge label={node[Above]{}}
    ]
    ]
  ] 
  [,edge label={node[Above]{$T_1$}}
    [,edge label={node[Below]{}}
    [,edge label={node[Below]{}}
    ]
    [,edge label={node[Above]{}}
    ]    
    ]
    [,edge label={node[Below]{}}
    [,edge label={node[Below]{}}
    ]
    [,label=Class C,edge label={node[Above]{}}
    ]    
    ]
  ]
]
\end{forest}
\caption{Illustration of Learning in Random Forest: Firstly, using raw data $D$, a training dataset is created. From this, a set of $m$ samples is randomly selected, and using the bootstrap method, several hundreds of decision trees, $\{T_1, T_2, \ldots, T_N\}$ are spawned (such as the one highlighted in the dotted box). At each node of these trees, only a random subset of features is considered. For a given test dataset (for example a new time series with an unknown class), a prediction is made by each decision (for example, here $T_1$ predicts Class C). A majority voting of these classes is made which results in the final prediction (here, Class C).}
\label{fig:decision_tree}
\end{figure}
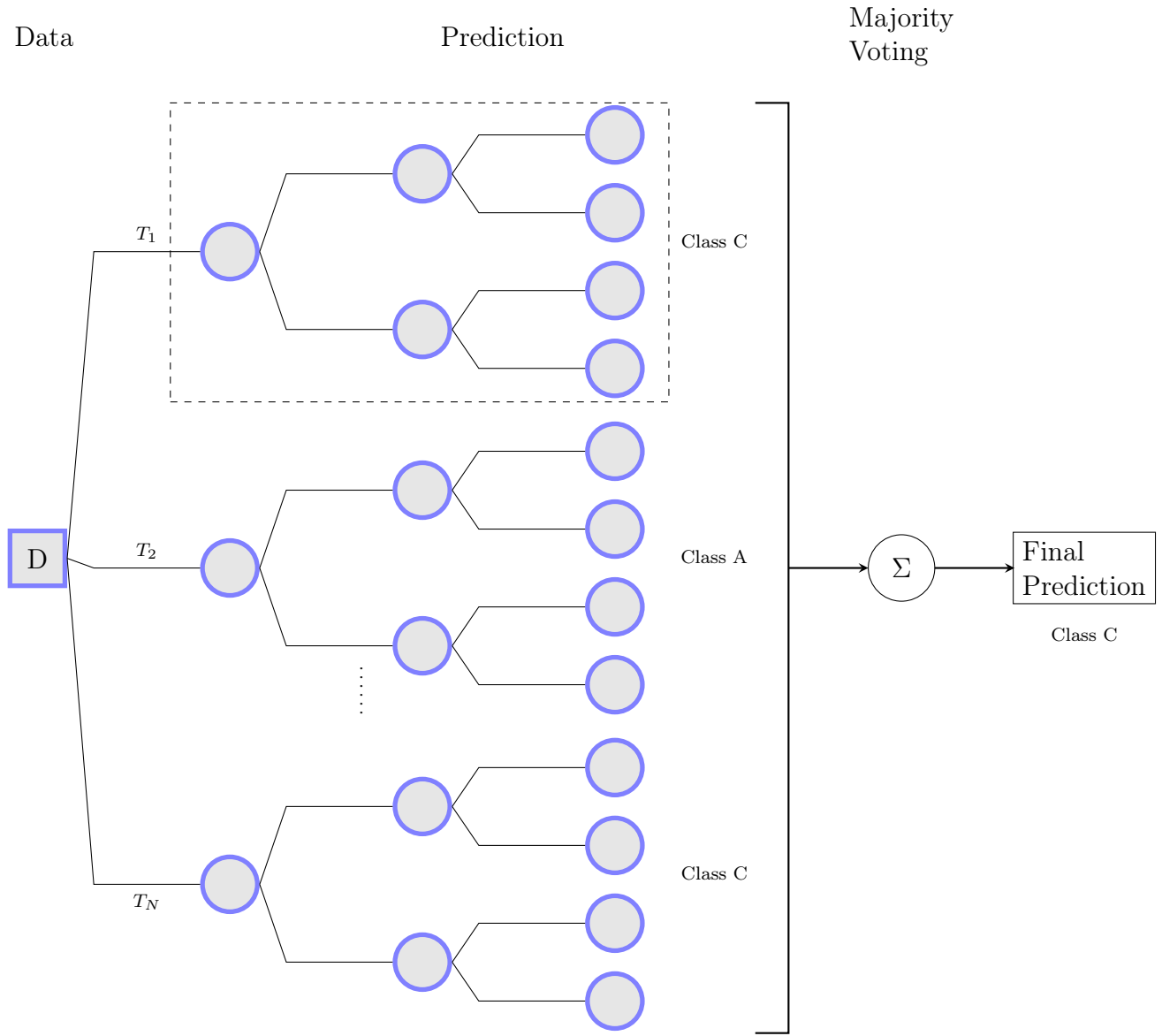

The decision tree-based classifiers emerged as the top family of models for all amperometric candidate datasets analyzed in this work. The decision tree is a data mining technique that predicts the value of a target variable based on a set of input features. A single decision tree has high variance since it may often run into the problem of overfitting as the tree grows deeper. This is because it learns from only one pathway of decisions (such as the dotted box in \fig\ \ref{fig:decision_tree}) and hence, may not make accurate predictions on unseen data. To circumvent this problem, ensemble methods including Random Forest and Extra Trees have been widely used. A Random Forest classifier is composed of many such individual trees that make independent model predictions. Majority voting is one of the most popular methods used in which each decision tree votes for a specific class and the ultimate predicted class is the one that has the maximum number of votes (such as that illustrated in \fig\ \ref{fig:decision_tree}). The class with the most votes becomes the final model prediction. As a result of this, predictions of individual decision trees do not affect that of others and hence have low correlations, thereby making the model robust. 

%%%%%%%%%%%%%%%%%%%%%%%%%%%%%%%%%%%%%%%%%%%%%%%%%%%%%%%%%%%%%%%%%%%%
\subsection*{Bootstrapping and Splitting}
\label{subsec:bootstrapping}

Random Forest classifiers have lower variance than decision trees since it reduces the risk of overfitting by introducing randomness. This is done by building multiple trees, randomly drawing observations with replacement (bootstrapping), splitting nodes on the best split among a random subset of the features selected at every node. Bootstrapping is a resampling method that is used to estimate statistics on a given population by sampling a dataset with replacement and by averaging estimates from multiple small data samples. This is a very useful method that is used very commonly in machine learning models used to make predictions on data that is not included in the training data. 
Splitting is a process of converting non-homogeneous parent nodes into the best possible two homogeneous child nodes. Hence the Random Forest classifiers reduce overfitting by building multiple trees and by splitting nodes on the best split. Extra Trees or Extremely Randomized Trees is similar to the Random Forest, in that it builds multiple trees and splits nodes using random subsets of features, however with two key differences. It does not bootstrap observations \ie\ it samples without replacement, and nodes are split on random splits among a random subset of features selected at every node. 

%%%%%%%%%%%%%%%%%%%%%%%%%%%%%%%%%%%%%%%%%%%%%%%%%%%%%%%%%%%%%%%%%%%%
\subsection*{Random Forest Classifier}
\label{subsec:rfclass}

Random Forest is a bagging method that creates multiple datasets by sampling with replacement from the training set, which is then used to train a classifier. However, Extra Trees is another bagging method that uses the whole original sample. On the other hand, an alternative method called boosting exists in which the learning algorithm uses a different weighting of the training set at each iteration. XGBoost or Extreme Gradient Boosting is an implementation of gradient boosted decision trees for speed and performance. It belongs to a family of boosting algorithms \ie\ combines a set of weak learners and delivers improved prediction accuracy, and uses the Gradient Boosting (GBM) framework. An overview of these decision tree classifier variants is summarized in \tab\ \ref{tab:trees_comp}. For the Random Forest Classifier, we used $20$ decision trees per forest with a total number of forests up to $500$. Hyperparameter values used for the other models were $20$ decision trees in Extra Trees and we used \enquote{gbtree} booster with learning rate, $\eta=0.3$ and minimum split loss parameter, $\gamma=0$ for XGBoost. It is important to note that while the systematic approach would be to do a thorough hyperparameter optimization, a user with limited time or computational resources may also do a limited hyperparameter search to train their model.

\begin{figure}[!htbp]
(A)\includegraphics[scale=0.4]{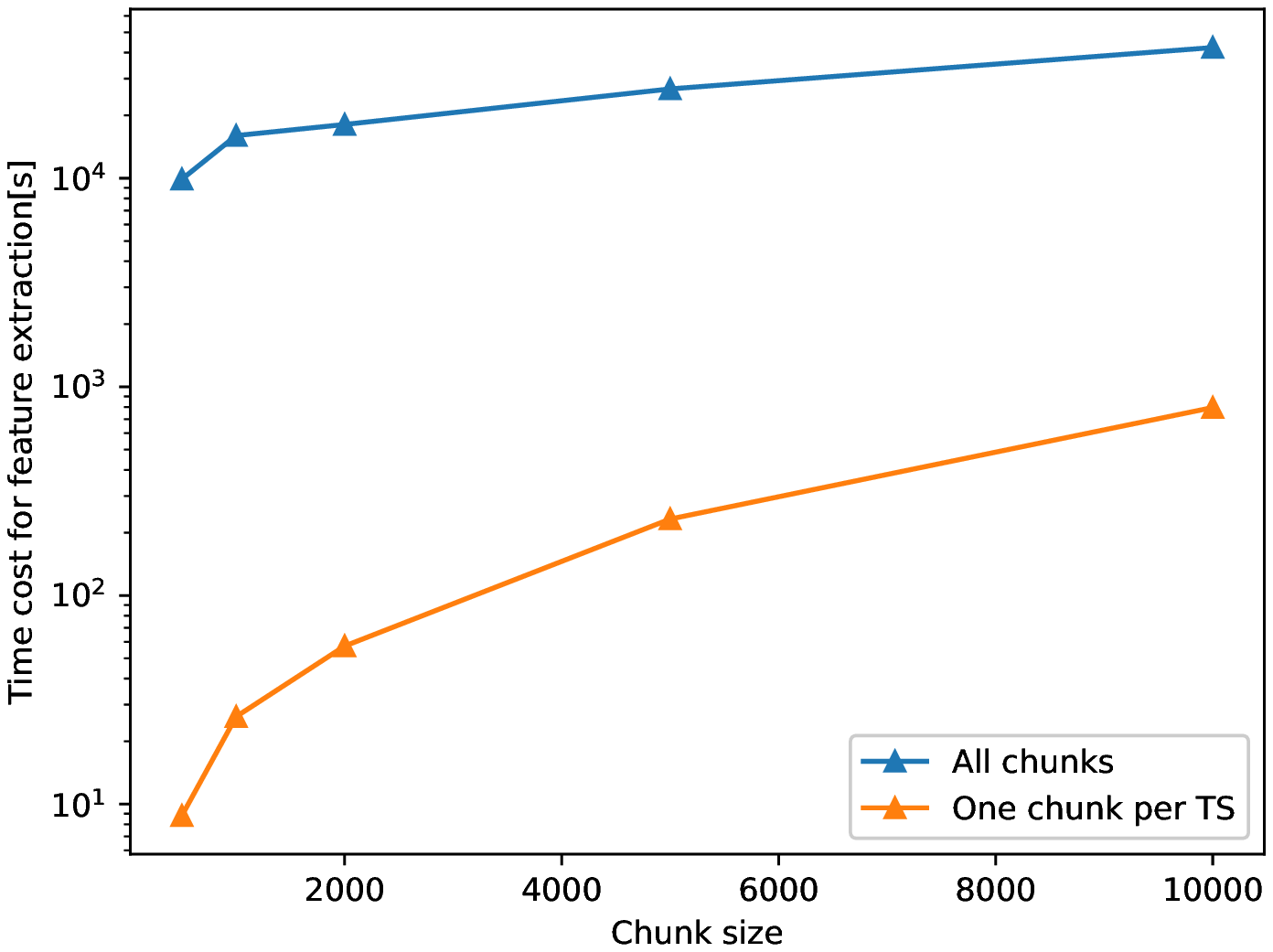}
(B)\includegraphics[scale=0.4]{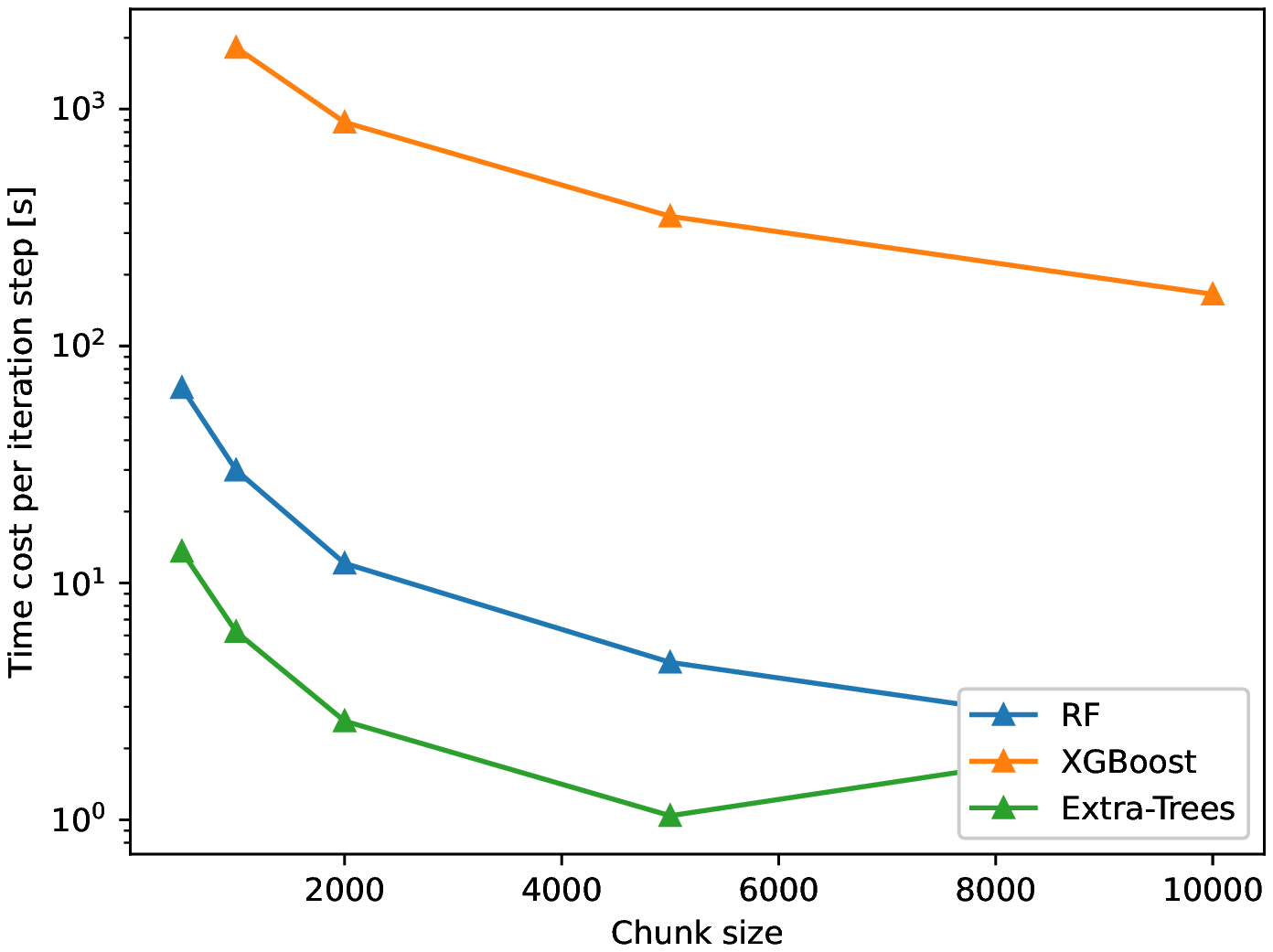}
\caption{Computational cost of the full method: (A) Feature extraction: Time cost for feature extraction from all chunks or one chunk per time series, given a chunk size (B) Classification: Time cost per iteration step for a given chunk size for all three Decision Tree classifiers.}
\label{fig:cost}
\end{figure}

All computational analyses were performed in Python. To get an idea of the typical timescale for the proposed workflow, consider that when running the feature extraction on cluster it took about $12$ hours for all chunked data given a chunk size of one second. Note that runtime was evaluated by running on one node ($24$ threads) on the RWTH-CLAIX HPC-Cluster. However, the classification itself often took only less than a few hundred seconds up to several minutes depending on the classifier. As can be expected, the time cost of feature extraction using \tsf\ increases exponentially with increasing chunk sizes, whereas using one chunk per time series takes a significantly shorter time than using full chunks as can be seen in \fig\ \ref{fig:cost}(A). The Extra Trees classifier is almost twice as fast as the Random Forest classifier, XGBoost has the highest computational cost per iteration for all chunk sizes as can be seen in \fig\ \ref{fig:cost}(B). A lean version of the implementation of these models is available as a Python package (more information in the supplementary section).
%%%%%%%%%%%%%%%%%%%%%%%%%%%%%%%%%%%%%%%%%%%%%%%%%%%%%%%%%%%%%%%%%%%%
\subsection*{Prediction Accuracy}
\label{subsec:accuracy}
Using this trained classifier, we can proceed to predictions on the test dataset. We will use the \hsd\ as the primary example to demonstrate the workflow and then discuss extensions to other datasets. In the \hsd, we used $593$ filtered features from \tsf\ with a train-to-test split ratio of $80/20$. This resulted in a training feature shape of $(7738, 593)$ and testing feature shape of $(1935, 593)$ (second row of \tab\ \ref{tab:shape_acc}). By using just the first $10000$ points or the beginning one second from each time series of the \hsd\ (even though the first second barely includes any amperometric spike events), we could already achieve $86\%$ accuracy. This suggests that fixing the bin position increases the accuracy to a certain extent, because the first second of data almost always contains baseline information, thereby facilitating the classification of time series from two different classes. Randomly slicing out a chunk performs worse than in the case of prediction on a fixed time window (dotted black curve) since it loses this baseline information.

\begin{figure}[!htbp]
\includegraphics[scale=0.5]{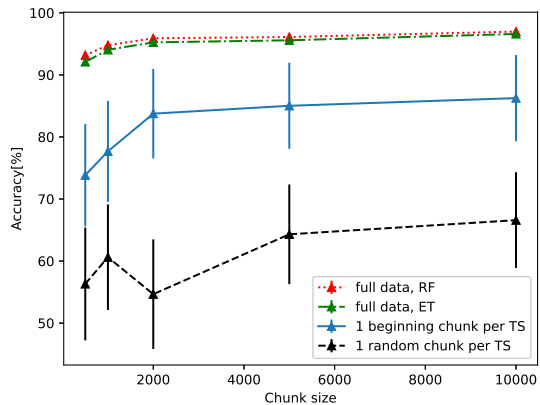}
\caption{Comparison of Random Forest and ExtraTrees Accuracy and standard deviation for different chunk sizes of the Hofmeister Series Dataset. Highest standard deviation for Random Forest with full data is $0.53\%$.}
\label{fig:accuracy}
\end{figure}

The accuracy of prediction is higher for smaller chunk sizes and lower for larger chunk sizes as shown in \fig\ \ref{fig:accuracy}. The model that uses one chunk per time series has a standard deviation of $6.9 - 8.2\%$ which decreases slightly as the chunk size increases (blue curve). For the full data model, the highest standard deviation is $\approx 0.53\%$, proving that a high accuracy holds when high-quality features are used. The slope of one chunk per time series is asymptotically similar to that of the fully chunked dataset. Using the full dataset gives as high as $97\%$ accuracy, as can be seen from the red dotted line (and second row of \tab\ \ref{tab:shape_acc}). The standard deviation of $0.41\%$ was calculated by computing averages over multiple realizations of Random Forest models. In addition, the prediction accuracy of Random Forest and Extra Trees are within a standard deviation of each other for all chunk sizes (green and red dotted curves). While the speed of XGBoost was not significantly better than the Random Forest classifier, the accuracy was similar on the \hsd. For a different number of forests, $N$, similar significant features are observed. For an increasing number of forests, the similarity in the top features becomes higher. The importance value was also similar for top features. The distribution of importance values was relatively thinly spread among the features extracted from \tsf\ and there was no single feature with a very high importance value. In other words, most top features have importance values $\geq 0.01$. A detailed treatment of the top one percent of the features for the \hsd\ and their inferences are discussed in detail in the results and discussion section. 

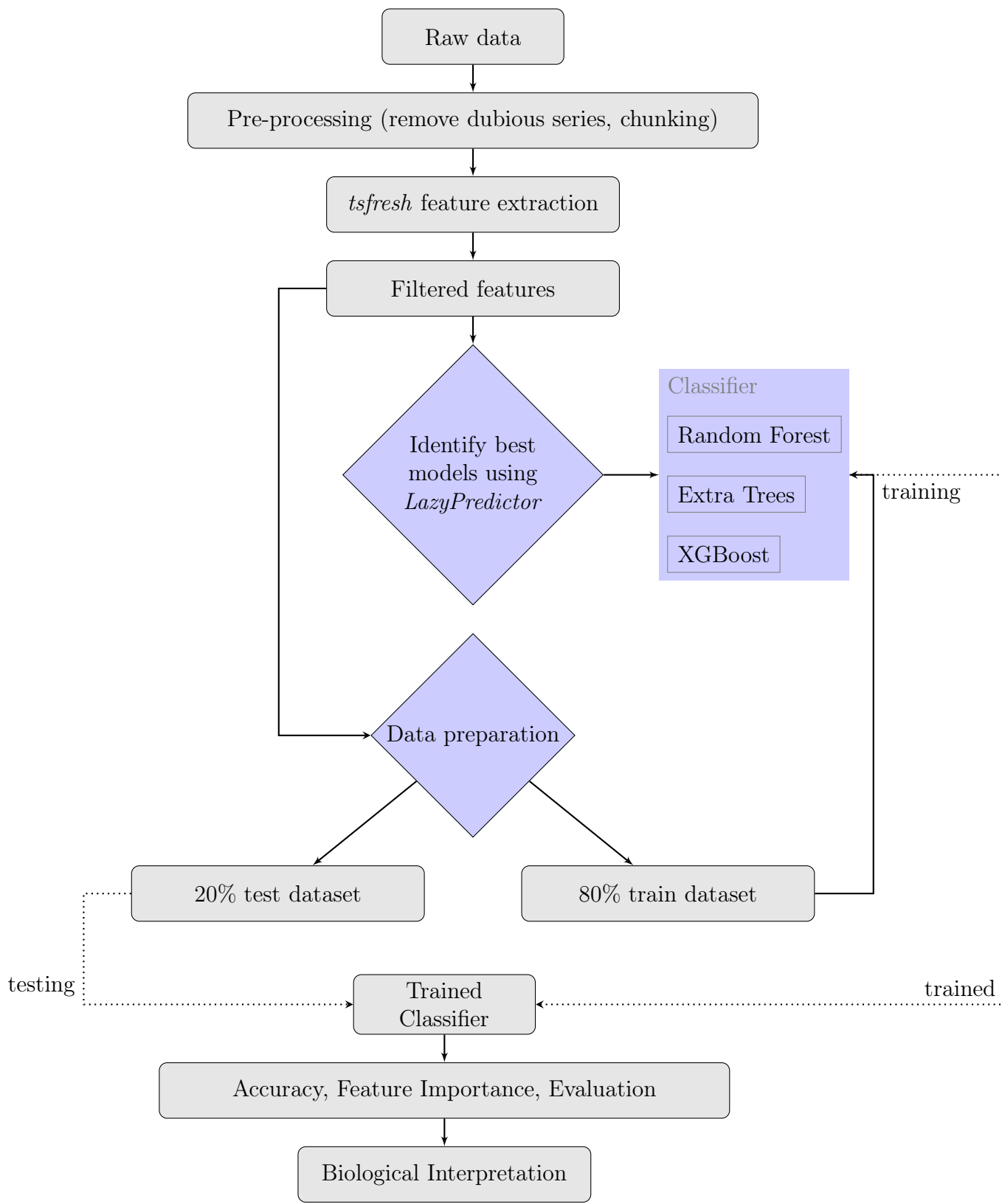
\begin{figure}[!htbp]
\begin{tikzpicture}[node distance = 3.0cm, auto, 
  mymatrix/.style={matrix of nodes, nodes=typetag, row sep=1em, fill=blue!20},
    mycontainer/.style={draw=none, inner sep=1ex},
  typetag/.style={draw=gray, inner sep=1ex, anchor=west},
  title/.style={draw=none, color=gray, inner sep=0pt}]
  \tikzstyle{decision} = [diamond, draw, fill=blue!20,
    text width=8em, text badly centered, node distance=1cm, inner sep=0pt]
  \tikzstyle{block} = [rectangle, draw, fill=gray!20,
    text width=5cm, text centered, rounded corners, minimum height=2cm]
  \tikzstyle{sblock} = [rectangle, draw, fill=gray!20,
    text width=3cm, text centered, rounded corners, minimum height=1cm]
  \tikzstyle{lblock} = [rectangle, draw, fill=gray!20,
    text width=10cm, text centered, rounded corners, minimum height=1cm]
  \tikzstyle{mblock} = [rectangle, draw, fill=gray!20,
    text width=5cm, text centered, rounded corners, minimum height=1cm]
  \tikzstyle{line} = [draw, -latex']  
  \tikzstyle{cloud} = [draw, ellipse,node distance=3cm,
    minimum height=1em, text width=5em]
	\tikzstyle{arrow}=[thick,->,>=stealth]
   
    \node[sblock] (raw) {Raw data};
    \node[lblock, below=0.5cm of raw] (pre) {Pre-processing (remove dubious series, chunking)};
    \node[mblock, below=0.5cm of pre, node distance=3cm](tsf) {\textit{tsfresh} feature extraction};
    \node[mblock, below=0.5cm of tsf] (ff) {Filtered features};
    \node[decision, below=0.5cm of ff] (laz) {Identify best models using \laz};
    \node[decision, below=0.5cm of laz] (prep) {Data preparation};
    \node[mblock, below=0.5cm of prep, xshift=-100] (test) {$20\%$ test dataset};
    \node[mblock, below=0.5cm of prep, xshift=100] (train) {$80\%$ train dataset};
  \matrix[mymatrix, right=1cm of laz] (cls) {
    |[title]|Classifier \\
    Random Forest \\
    Extra Trees\\
    XGBoost \\
  };
    \node[sblock, below=2cm of test, right of= test] (trained) {Trained Classifier};
    \node[lblock, below=0.5cm of trained] (metrics) {Accuracy, Feature Importance, Evaluation};
    \node[mblock, below=0.5cm of metrics] (interp) {Biological Interpretation};

   \node[mycontainer, fit=(cls)] {};

    % arrows
	\path [line, thick] (raw) -- (pre);
	\path [line, thick] (pre) -- (tsf);
	\path [line, thick] (tsf) -- (ff);
	\path [line, thick] (ff) -- (laz);
	\path [line, thick] (laz) -- (cls);
	% \path [line, thick] (laz) -- (prep);
	\path [line, thick] (prep) -- (test);
	\path [line, thick] (prep) -- (train);
	\path [line, thick] (trained) -- (metrics);
	\path [line, thick] (metrics) -- (interp);

	% special arrows
	\draw [arrow] (ff) --   ++(-3.5,0) |- (prep);
	\draw [arrow, dotted] (test)  --   ++(-3.5,0) |- node [left, anchor = south east] {testing} (trained);
	\draw [arrow] (train) --   ++(3.7,0) |- node [left, anchor = north west] {training} (cls);
	\draw [arrow, dotted] (cls) --   ++(4.5,0) |- node [left, anchor = south east] {trained} (trained);

\end{tikzpicture}
\caption{Illustration of workflow adopted for amperometric time series classification.}
\label{fig:workflow}
\end{figure}

The electrodes dataset is an VIEC experiment dataset conducted by insertion of different types of micro-electrodes including Carbon, Platinum, and Gold. The electrodes dataset has fewer measurements per class as compared to the \hsd. However the measurements per experiment were significantly longer (of the order of $10^3$ seconds), which helped provide similar unfiltered feature sizes for training, \ie\ $(10122, 495)$ and a testing set of $(2531, 495)$ (second row of \tab\ \ref{tab:shape_acc}). After filtering, $416$ features were found to be important, and with this, the Random Forest Classifier achieved an accuracy of $\approx 98\%$. The cells dataset is data-rich and the resultant feature matrix for training and test set was sufficient to achieve an accuracy of $99.714\%$ with a very low standard deviation of $0.102\%$ for all models (as shown in \tab\ \ref{tab:shape_acc}). Extra Trees showed similar accuracy of $99.559\%$ and a standard deviation of $0.111\%$. XGBoost for the cells dataset showed an accuracy of $99.835\%$ and a standard deviation of $0.059\%$. Other tree-based classifiers such as Extra Trees and XGBoost give comparable results $(\pm 0.12\%)$. Similar standards of accuracy were also observed for these models for the VIEC dataset (as shown in \tab\ \ref{tab:shape_acc}). The classification workflow shown in \fig\ \ref{fig:workflow} has therefore been validated for multiple datasets using different variants of decision tree models with consistently high accuracy, demonstrating the universal applicability of the proposed classification workflow. The final step in the proposed workflow, which comes after achieving high prediction accuracies from a trained classifier, is to evaluate and understand the top features that rendered a good classifier. This aspect of feature interpretation is discussed in detail in the results and discussion section.
\clearpage

\clearpage

\clearpage

\begin{table}[!htbp]
\centering
\begin{tabular}{ |c|c|c|c| } 
 \hline
  & \textbf{Decision Tree} & \textbf{Random Forest} & \textbf{Extra Trees}\\ \hline \hline
 Number of trees & $1$ & Many & Many \\ \hline
 Features considered & All & Random & Random\\ \hline
 Bootstrapping & No & Yes & No \\ \hline
 How split is made & Best split & Best split & Random split\\\hline
\end{tabular}
\captionof{table}{Comparison of Tree-Based Learning Methods.}
\label{tab:trees_comp}
\end{table}

\begin{table}
\centering
\begin{tabular}{ |p{2cm}|p{2cm}|p{2cm}|p{2cm}|p{2cm}|p{2cm}|p{2cm}|} 
 \hline
 \textbf{Dataset} & \textbf{Feature source} & \textbf{Training Feature Shape} & \textbf{Testing Feature Shape} & \textbf{Accuracy (Random Forest), \%} & \textbf{Accuracy (Extra Trees) \%} & \textbf{Accuracy (XGBoost), \%}\\
 \hline \hline
 \rowcolor{lightgray}
 Hofmeister (\textit{QuantaAnalysis})& Only spikes & $(113, 10)$ & $(13, 10)$ & $26.15 \pm 4$ & $22.48\pm 3.75$ & $29.55\pm 2.64$\\
 \hline \hline
 \rowcolor{blue!20}
 Hofmeister Dataset (\tsf) & Full Time Series & $(7738, 593)$ &  $(1935, 593)$ & $96.30 \pm 0.41$ & $95.92 \pm 0.47$ & $98.34\pm0.18$\\
 \hline
 \rowcolor{blue!20}
 Electrodes Dataset (\tsf) & Full Time Series & $(10122, 495)$ & $(2531, 495)$ & $98.37 \pm 0.28$ & $98.07 \pm 0.33$ & $99.12 \pm 0.07$\\
 \hline
 \rowcolor{blue!20}
 Cells Dataset (\tsf)& Full Time Series & $(12093, 466)$ & $(3024, 466)$	& $99.71 \pm 0.70$ & $99.60 \pm 0.11$ & $99.84 \pm 0.059$\\
 \hline
 \rowcolor{blue!20}
 VIEC Dataset  (\tsf)& Full Time Series & $(11532, 570)$ & $(2883, 570)$ 	& $99.92 \pm 0.06$ & $99.93 \pm 0.06$ & $99.95\pm 0.02$\\
 \hline
\end{tabular}
\caption{Overview of training and testing feature shape and corresponding accuracy $\pm$ standard deviation of different decision tree classifiers for the candidate datasets. The first row shows the feature shape for the Hofmeister Series dataset extracted using \textit{QuantaAnalysis} in IgorPro (gray cells). Subsequent rows show the shape of features extracted from \tsf\ (blue cells).}
\label{tab:shape_acc}
\end{table}
%%%%%%%%%%%%%%%%%%%%%%%%%%%%%%%%%%%%%%%%%%%%%%%%%%%%%%%%%%%%%%%%%%%%%%%%%%%%%%%%%%%%%%%%%%%%%%%%%%%%%%%%%%%%%%%%%%%%%%%%%%%%%%%%%%%%%%%%%%%%%%%%%%%%%%%%%%%%%%%%%%%%%%%%%%%%%%%%%%%%%%%%%%%%%%%%%%%%%%%%%%%%%%%%%%%%%%%%%%

%%%%%%%%%%%%%%%%%%%%%%%%%%%%%%%%%%%%%%%%%%%%%%%%%%%%%%%%%%%%%%%%%%%%%
%% The "Acknowledgement" section can be given in all manuscript
%% classes.  This should be given within the "acknowledgement"
%% environment, which will make the correct section or running title.
%%%%%%%%%%%%%%%%%%%%%%%%%%%%%%%%%%%%%%%%%%%%%%%%%%%%%%%%%%%%%%%%%%%%%
%%%%%%%%%%%%%%%%%%%%%%%%%%%%%%%%%%%%%%%%%%%%%%%%%%%%%%%%%%%%%%%%%%%%%
%%%%%%%%%%%%%%%%%%%%%%%%%%%%%%%%%%%%%%%%%%%%%%%%%%%%%%%%%%%%%%%%%%%%%
\section*{Funding}

This work was supported by the JPND Neuronode grant no: 01ED1802 for the computational work; and  the MSCA grant funding from the European Union's Horizon 2020 research and innovation program under the Marie Skłodowska-Curie Grant Agreement No.793324 for the experimental work.
%%%%%%%%%%%%%%%%%%%%%%%%%%%%%%%%%%%%%%%%%%%%%%%%%%%%%%%%%%%%%%%%%%%%%
%%%%%%%%%%%%%%%%%%%%%%%%%%%%%%%%%%%%%%%%%%%%%%%%%%%%%%%%%%%%%%%%%%%%%
\begin{acknowledgement}
The authors thank Mohaddeseh Aref, Zahra Taleat, Alex S. Lima, Elias Ranjbari, Johan Dunevall for their data contributions.
\end{acknowledgement}
%%%%%%%%%%%%%%%%%%%%%%%%%%%%%%%%%%%%%%%%%%%%%%%%%%%%%%%%%%%%%%%%%%%%%
%%%%%%%%%%%%%%%%%%%%%%%%%%%%%%%%%%%%%%%%%%%%%%%%%%%%%%%%%%%%%%%%%%%%%
%% The same is true for Supporting Information, which should use the
%% suppinfo environment.
%%%%%%%%%%%%%%%%%%%%%%%%%%%%%%%%%%%%%%%%%%%%%%%%%%%%%%%%%%%%%%%%%%%%%
\clearpage
\begin{suppinfo}
\label{sec:supp}
%%%%%%%%%%%%%%%%%%%%%%%%%%%%%%%%%%%%%%%%%%%%%%%%%%%%%%%%%%%%%%%%%%%%%
\subsection{Experimental Data Generation}
\label{subsec:data}

% \jk{All: 
% \begin{itemize}
% \item Please check if the summary table here is accurate
% \item Please check if the summary of data generation is correct. 
% \item Could you please add data generation procedure for electrodes dataset, cells and IVIEC/VIEC dataset. I do not have a reference for this.
% \end{itemize}
% }

%%%%%%%%%%%%%%%%%%%%%%%%%%%%%%%%%%%%%%%%%%%%%%%%%%%%%%%%%%%%%%%%%%%%%
% -  Notes on experimental data generation: perhaps include text on how the data was generated? 
% -  Add link to the decision tree program (appendix may not be required for this paper?)
% -  Check appendix in the report (tsfresh, lazypredictor) @done
% -  Feature description @done
% -  Detailed mathematical description of the methods
% -  Expected total length of paper: 7-8 pages
%%%%%%%%%%%%%%%%%%%%%%%%%%%%%%%%%%%%%%%%%%%%%%%%%%%%%%%%%%%%%%%%%%%%%
\begin{sidewaystable}[!htbp]
\centering
\begin{tabular}{ |c|c|c|p{50mm}|p{30mm}| } 
 \hline
 Attribute & Hofmeister Series & Electrodes & Cells & VIEC\\ 
 \hline \hline
 Method & SCA & VIEC & IVIEC & VIEC \\ 
 \hline
 Cell type & Adrenal chromaffin cells & Chromaffin cells & Different cell types & Chromaffin cells\\ \hline
 Conditions & $K^{+}$ and anions stimulation & Dipping Method & NA & adding and dipping method \\ \hline
 Categories & Hofmeister Ions  & Au, Pt, C & Beta, Chromaffin, PC12 cells, \textit{Drosophila Melanogaster} & Adding/ Dipping\\ \hline
 Length (sec) & $41 - 126$ & $105 - 1459$ & $59 - 562$ & $81 - 1203$\\ \hline
 Sampling Frequency (kHz) & $10$ & $10$ & $10$ kHz, for fly dataset $20$ kHz & $10$\\ \hline
 \# Samples & $158$ & $22$ & $123$ & $44$\\ 
 \hline
\end{tabular}
\caption{Summary of attributes of candidate datasets}
\label{tab:meta}
\end{sidewaystable}

A summary of attributes of candidate datasets used in our analysis is given in Table \ref{tab:meta}. A short summary of the procedure adopted to generate these datasets is summarized from \cite{He2020, Majdi2017} in the following subsections:
%%%%%%%%%%%%%%%%%%%%%%%%%%%%%%%%%%%%%%%%%%%%%%%%%%%%%%%%%%%%%%%%%%%%%
\clearpage
\subsubsection*{Hofmeister Series Dataset}
\label{subsubsec:hof}

\begin{figure}[!htbp]
    \centering
    \begin{subfigure}[b]{0.3\textwidth}
        \centering
        \includegraphics[scale=0.3]{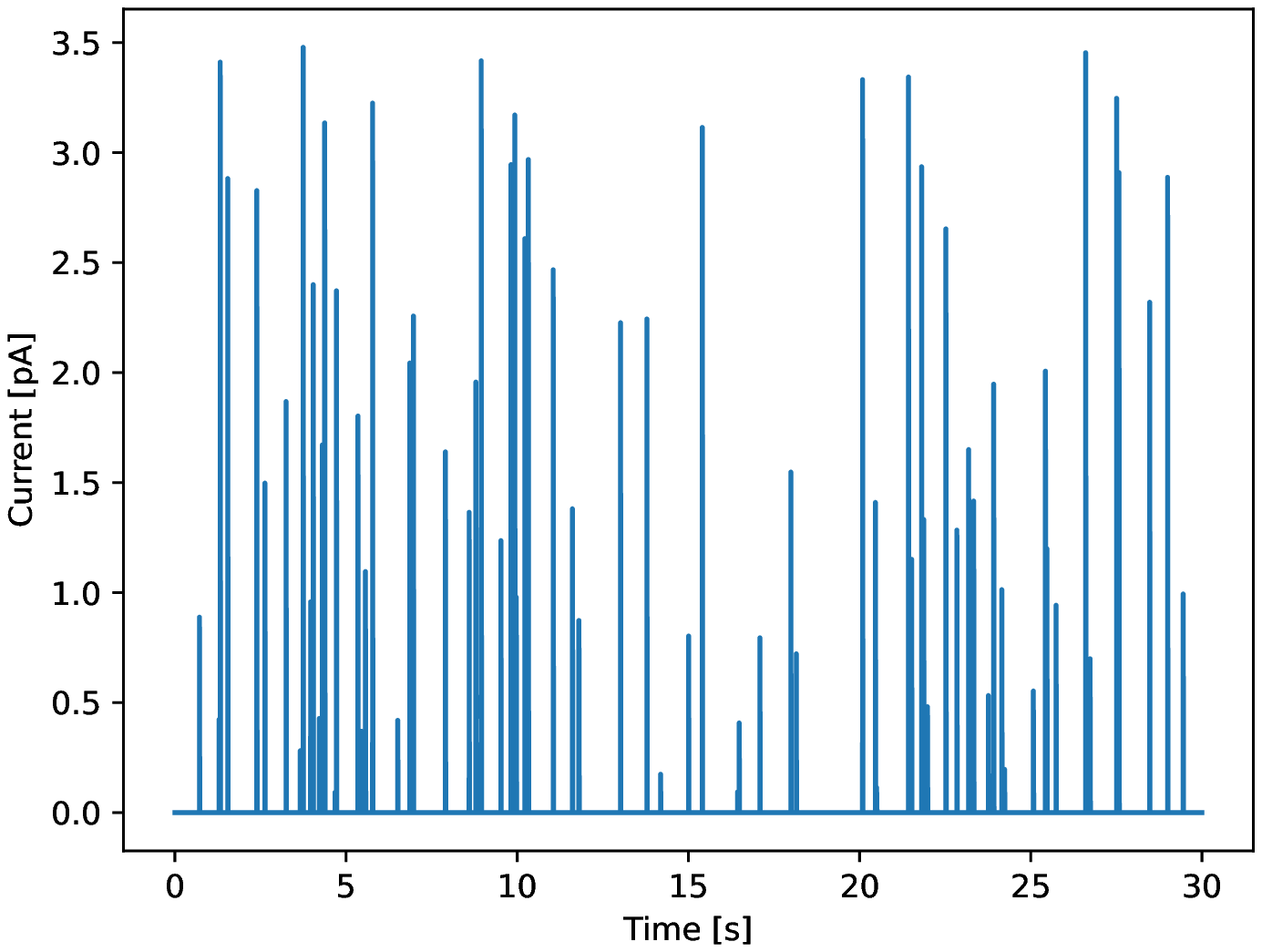}
        \caption[]%
        {{\small Spike trains with Br$^-$ stimulation}}    
        \label{fig:br}
    \end{subfigure}
    \begin{subfigure}[b]{0.3\textwidth}
        \centering
		\includegraphics[scale=0.3]{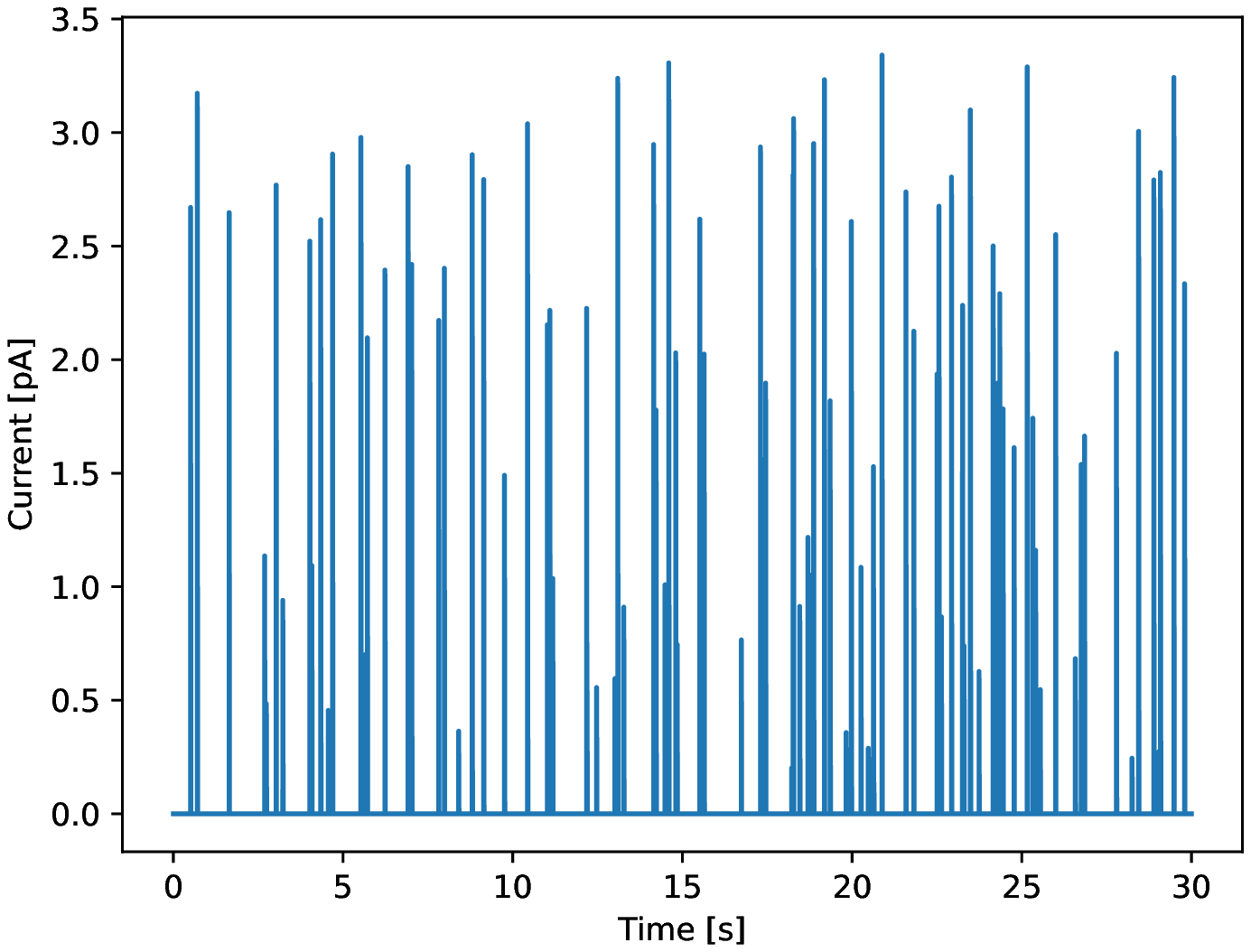}
        \caption[]%
        {{\small Spike trains with Cl$^{-}$ stimulation}}    
        \label{fig:cl}
    \end{subfigure}
    \begin{subfigure}[b]{0.3\textwidth}
        \centering
		\includegraphics[scale=0.3]{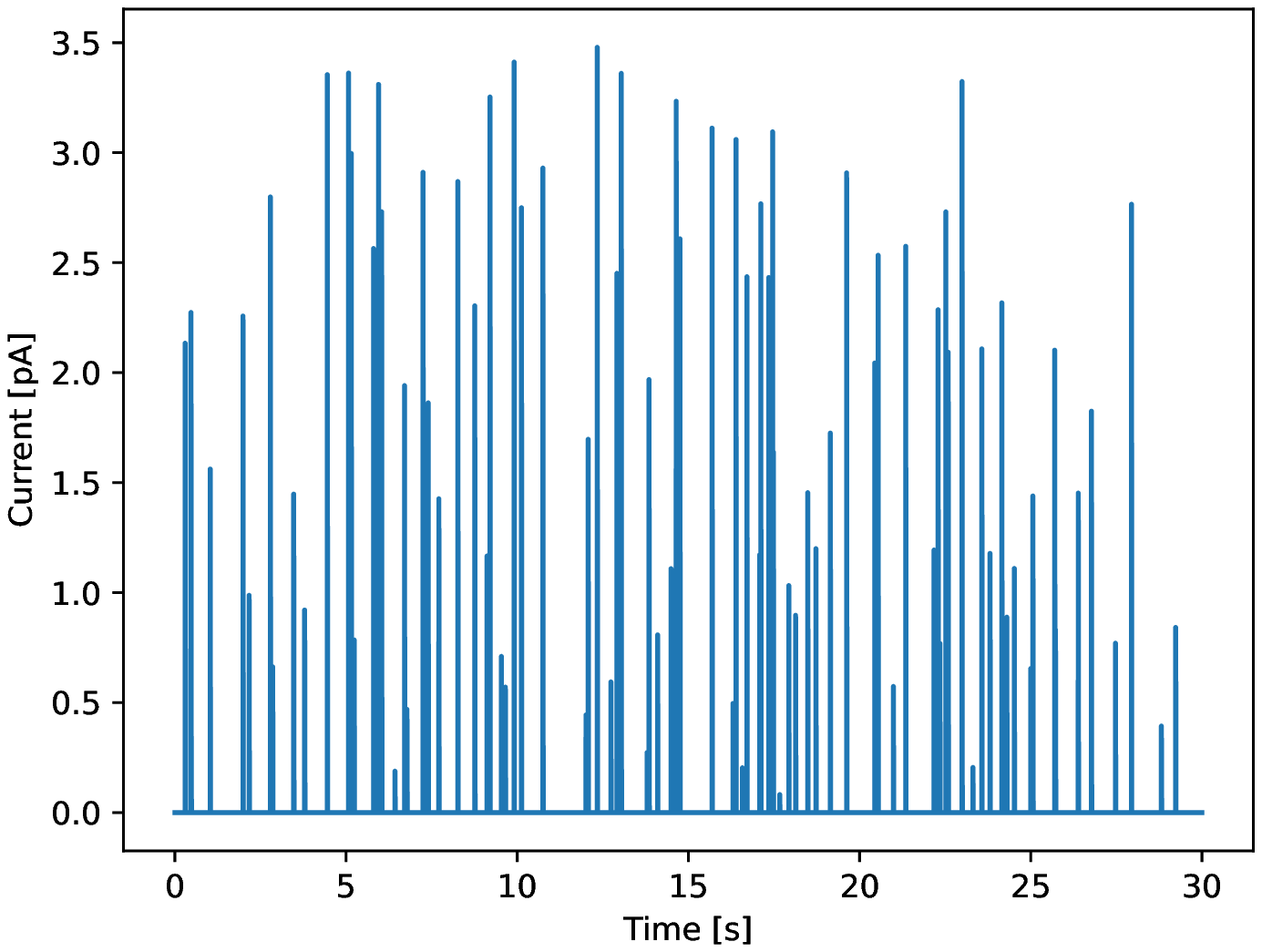}
        \caption[]%
        {{\small Spike trains with ClO$_{4}^{-}$ stimulation}}    
        \label{fig:clo4}
    \end{subfigure}
    \begin{subfigure}[b]{0.3\textwidth}
        \centering
		\includegraphics[scale=0.3]{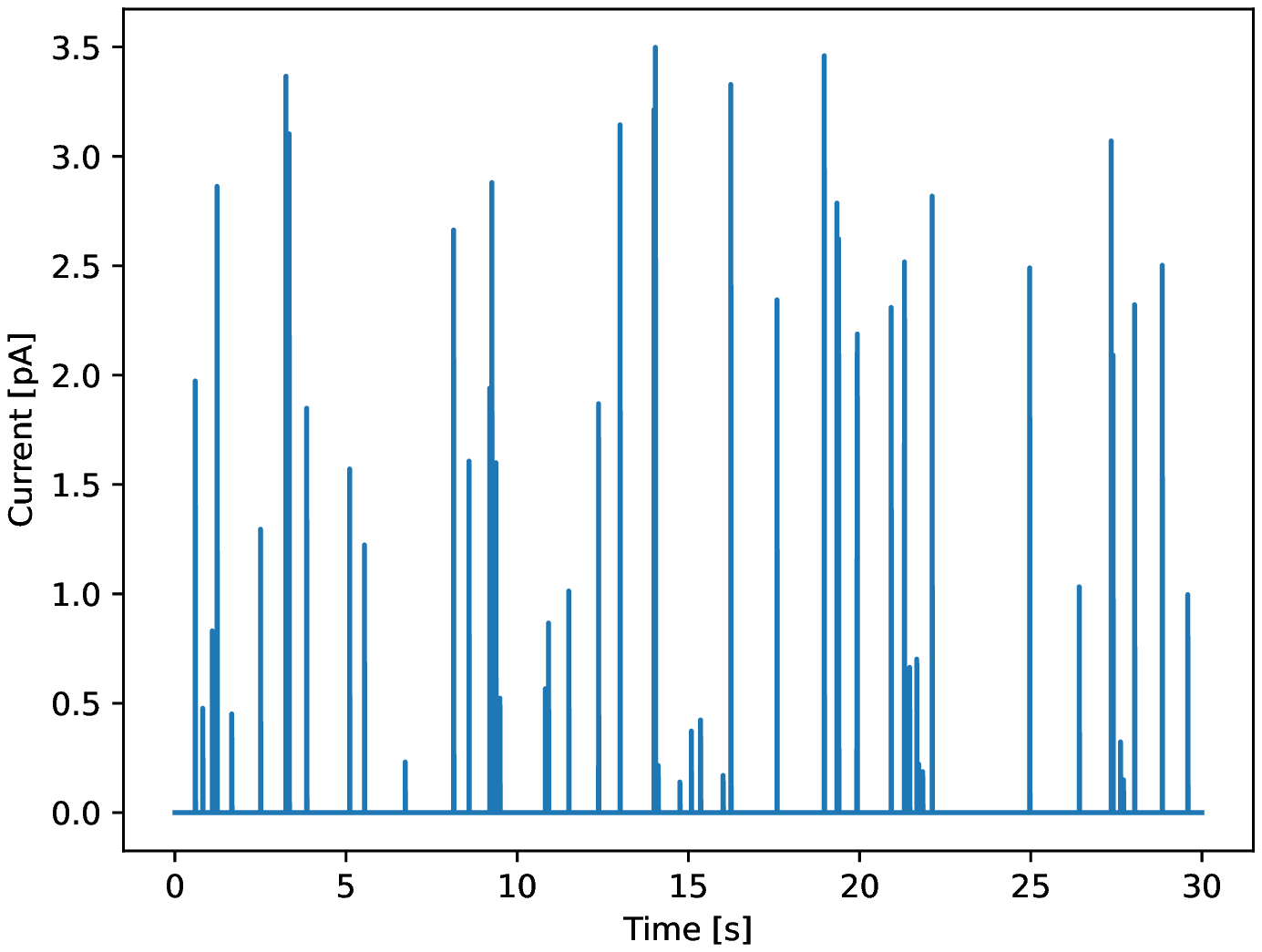}
        \caption[]%
        {{\small Spike trains with NO$_{3}^{-}$ stimulation}}    
        \label{fig:no3}
    \end{subfigure}
    \begin{subfigure}[b]{0.3\textwidth}
        \centering
		\includegraphics[scale=0.3]{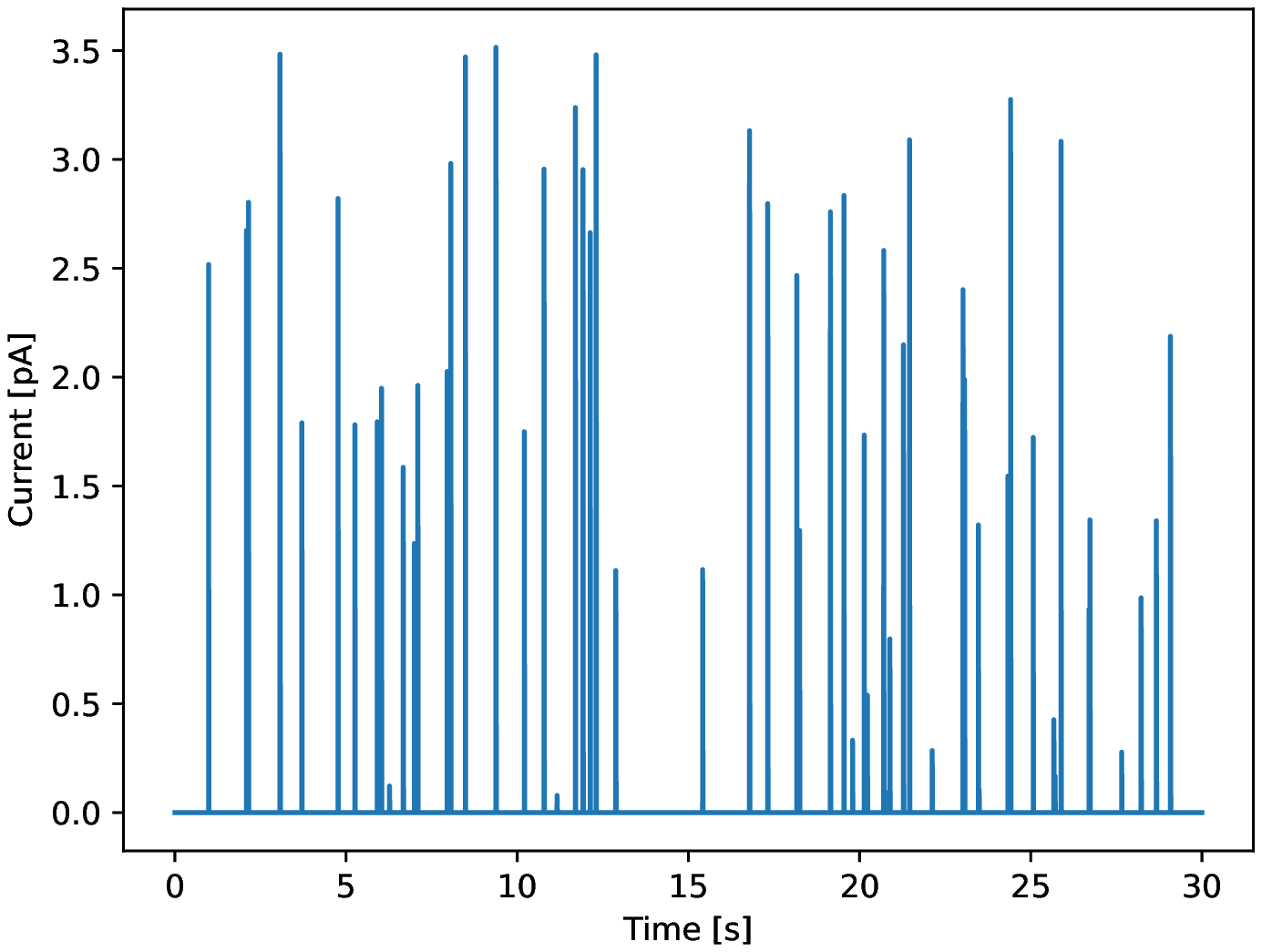}
        \caption[]%
        {{\small Spike trains with SCN$^{-}$ stimulation}}    
        \label{fig:scn}
    \end{subfigure}
    \caption{Amperometric traces of Chromaffin cells under different ion stimulations obtained through SCA experiments.}
    \label{fig:traces}
\end{figure}

Bovine adrenal glands were obtained from a local slaughterhouse and the cells were kept at $37$ degree Celsius in isotonic solution during the whole experimental process. Electrochemical recordings from single chromaffin cells were performed on an inverted microscope in a Faraday cage. The output was filtered at $2.1$ kHz and digitized at $5$ kHz. For single-cell exocytosis, the micro-disk electrode was moved slowly by a  patch-clamp micromanipulator to place it on the membrane of a chromaffin cell without causing any damage to the surface. Ten seconds after the start of recording, $30$ mM $K^{+}$ stimulating solution in a glass micropipette was injected into the surrounding of the chromaffin cells with a single $30$-s injection pulse.
%%%%%%%%%%%%%%%%%%%%%%%%%%%%%%%%%%%%%%%%%%%%%%%%%%%%%%%%%%%%%%%%%%%%%
\clearpage
\subsubsection*{Electrodes Dataset}
\label{subsubsec:electrodes}

\begin{figure}[!htbp]
\centering
\begin{subfigure}[b]{0.4\textwidth}
\includegraphics[scale=0.4]{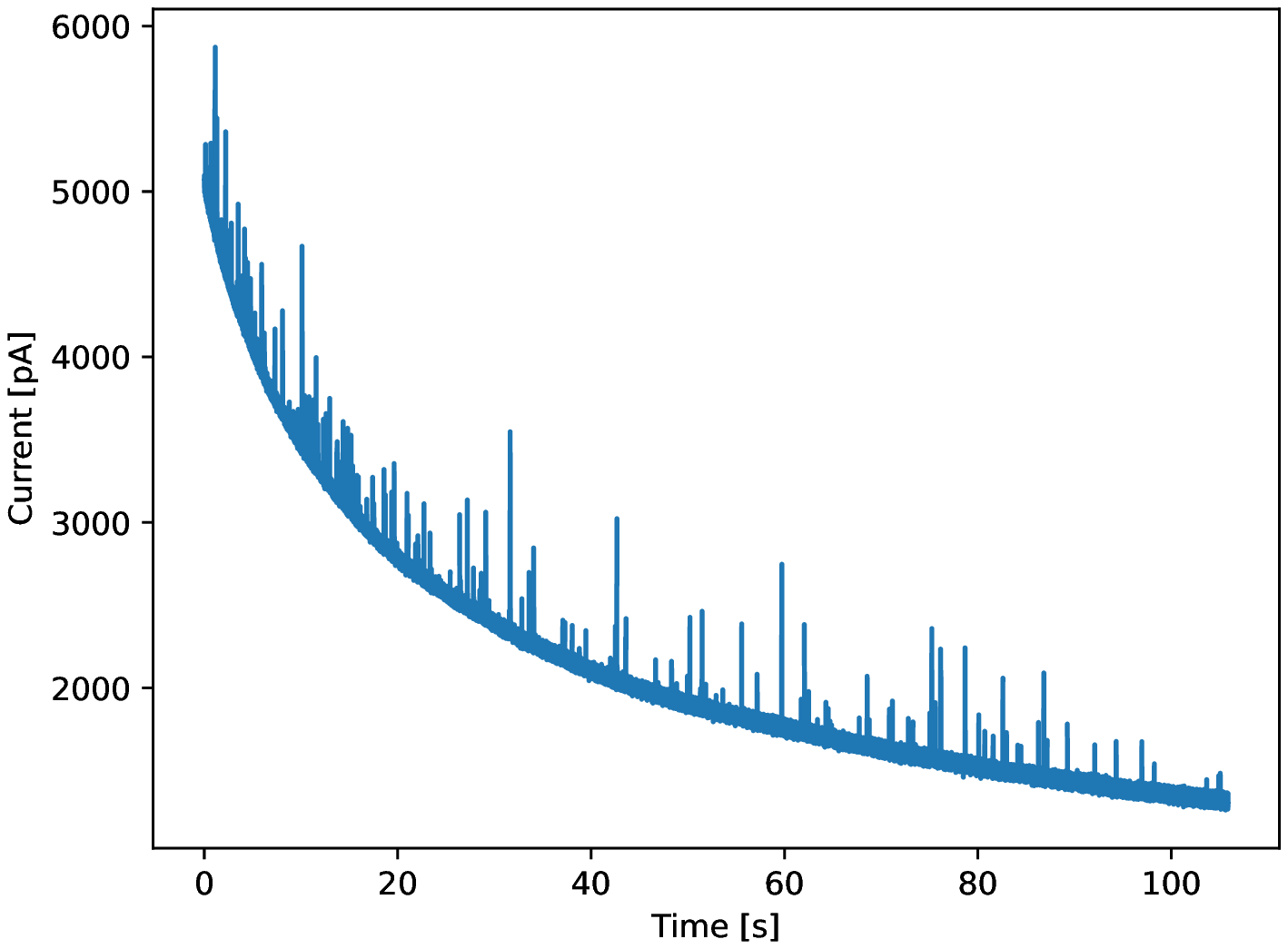}
\end{subfigure}
\begin{subfigure}[b]{0.4\textwidth}
\includegraphics[scale=0.4]{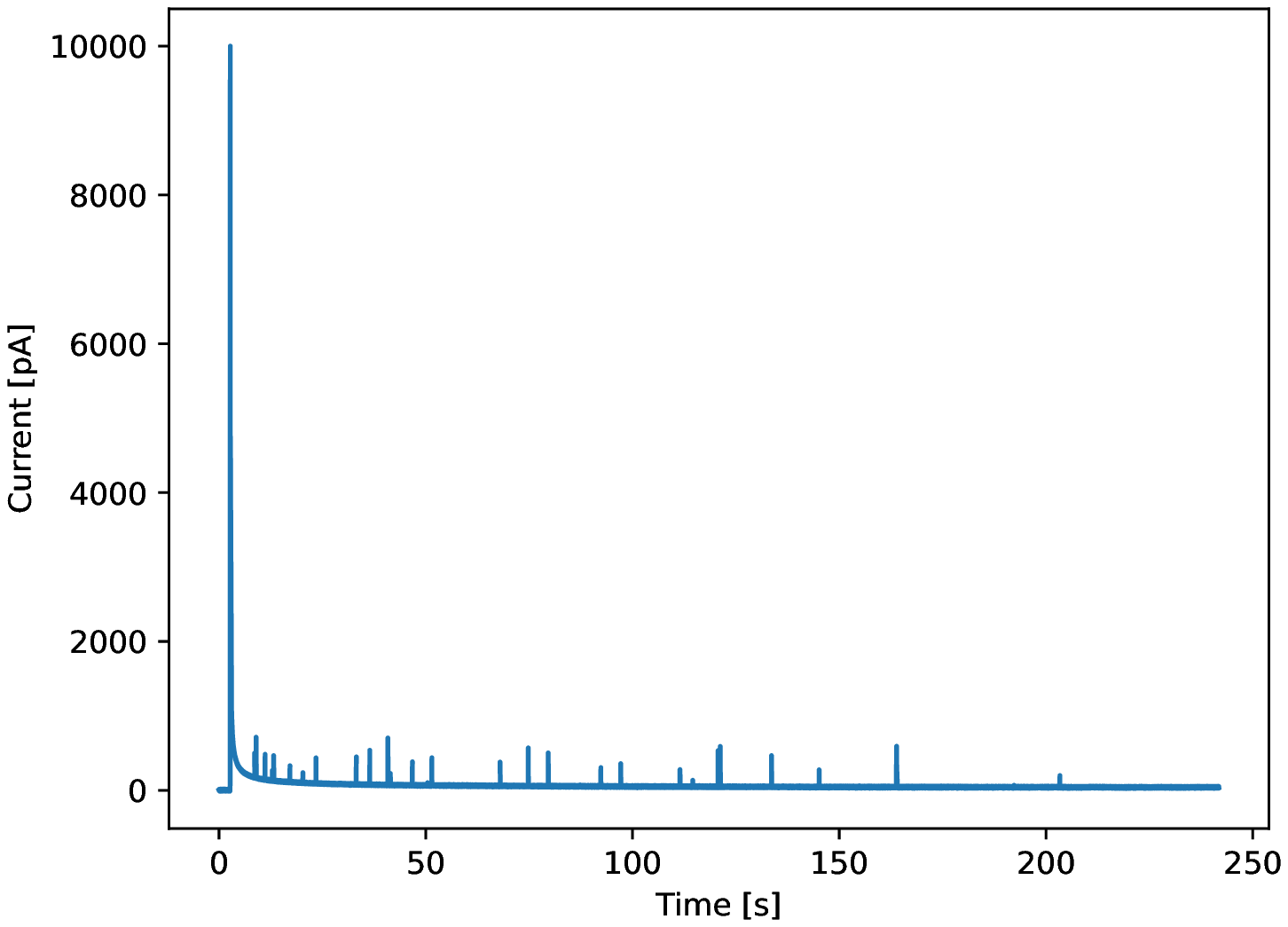}
\end{subfigure}\\
\begin{subfigure}[b]{0.4\textwidth}
\includegraphics[scale=0.4]{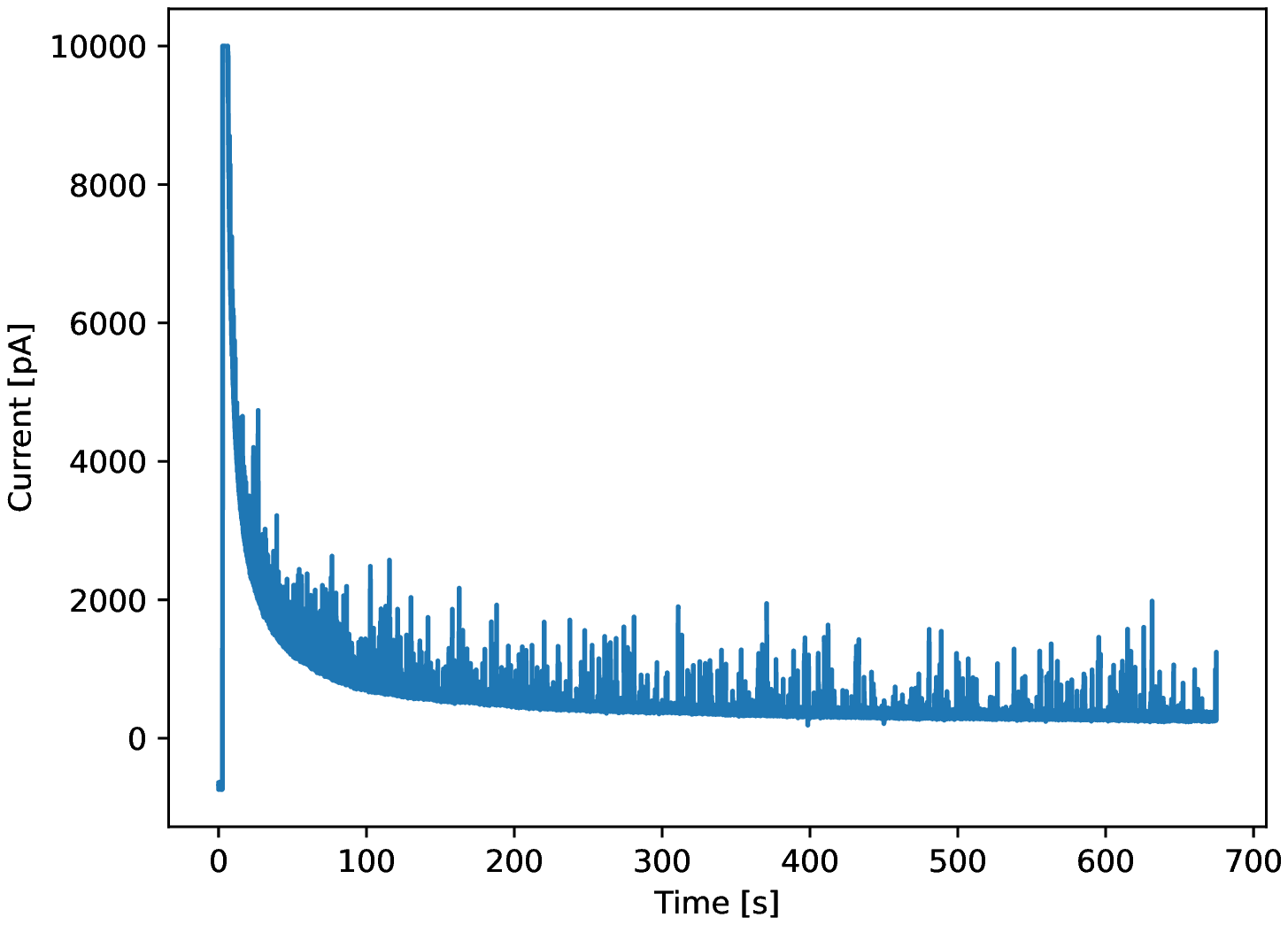}
\end{subfigure}
\caption{Amperometric traces of Chromaffin cells for different electrodes (gold, carbon and platinum).}
\end{figure}

Bovine adrenal glands were obtained from a local slaughterhouse and transported in cold Locke’s buffer. Glands were trimmed of surrounding fat and rinsed through the adrenal vein with Locke's solution. The medulla was detached from the cortex with a scalpel and then mechanically homogenized in ice-cold homogenizing buffer. The homogenate was centrifuged at $1000$ x g for $10$ minutes to eliminate non-lysed cells and cell debris. After that, the supernatant was subsequently centrifuged at $10000$ x g for $20$ min to pellet vesicles. All centrifugation was performed at $4^{\circ}$C. The final pellet of chromaffin vesicles was resuspended and diluted in homogenizing buffer for VIEC measurements.

For the VIEC experiments, the electrodes were first dipped in a vesicle suspension for $30$ minutes at $4^{\circ}$C and then placed in homogenizing buffer for $20$ minutes at $37^{\circ}$C for experimental recording. During the measurements, a constant potential of $+700$ mV vs. Ag/AgCl reference electrode was applied to the working electrode using a low current potentiostat (Axopatch $200$B, Molecular Devices, Sunnyvale, CA, U.S.A). The signal output was filtered at $2$ kHz using a $4$-pole Bessel filter and digitized at $10$ kHz using a Digidata model $1440$A and Axoscope $10.3$ software (Axon Instruments Inc., Sunnyvale, CA, U.S.A.). 

Briefly, a carbon fiber with $33\mu m$ diameter was aspirated into a borosilicate capillary ($1.2$ mm $O.$D., $0.69$ mm I.D., Sutter Instrument Co., Novato, CA, U.S.A.). The capillaries were subsequently pulled using a micropipette puller (Narishige Inc., London, U.K) and the carbon fiber was cut at the glass junction. The gap between the carbon fiber and glass was sealed by dipping the pulled tip in epoxy. The glued electrodes were placed in an oven at $100^{\circ}$C overnight to complete the sealing step. The sealed electrodes were beveled at $45^{\circ}$ angle (EG-$400$, Narishige Inc., London, U.K.). A similar procedure was utilized for gold and platinum disk microelectrode fabrication. Here, either a $1$-cm length of $125\mu m$ diameter Au wire or $100\mu m$ diameter Pt wire (Goodfellow, Cambridge Ltd. U.K.) that were connected to a longer piece of a conductive wire (silver wire, $10$ cm) using silver paste were inserted into the pulled capillary and similarly sealed with epoxy and beveled at a $45^{\circ}$ angle. 

%%%%%%%%%%%%%%%%%%%%%%%%%%%%%%%%%%%%%%%%%%%%%%%%%%%%%%%%%%%%%%%%%%%%%
\clearpage
\subsubsection*{Cells Dataset}
\label{subsubsec:cells}

\begin{figure}[!htbp]
\centering
\begin{subfigure}[b]{0.4\textwidth}
\includegraphics[scale=0.4]{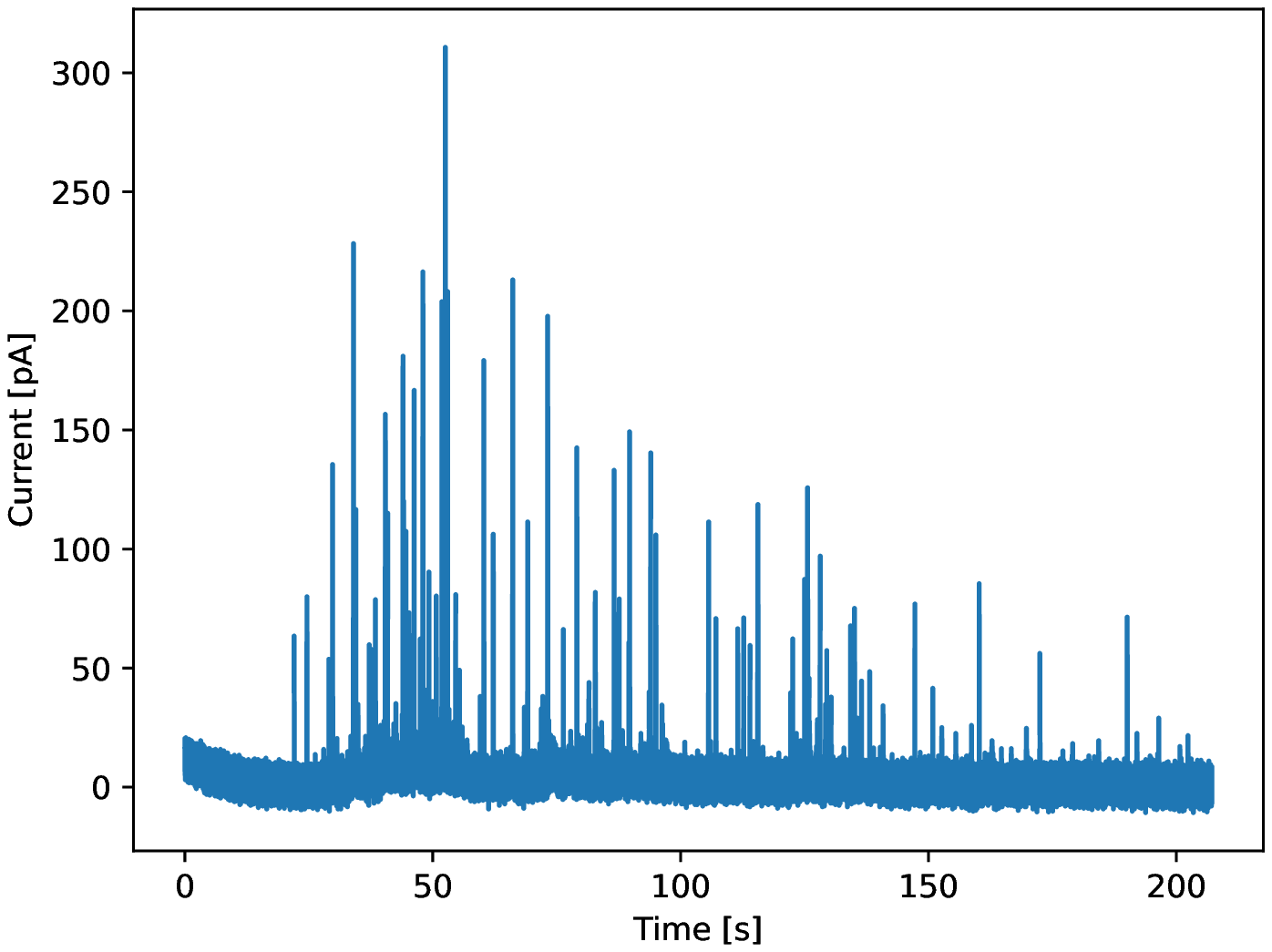}
\end{subfigure}
\begin{subfigure}[b]{0.4\textwidth}
\includegraphics[scale=0.4]{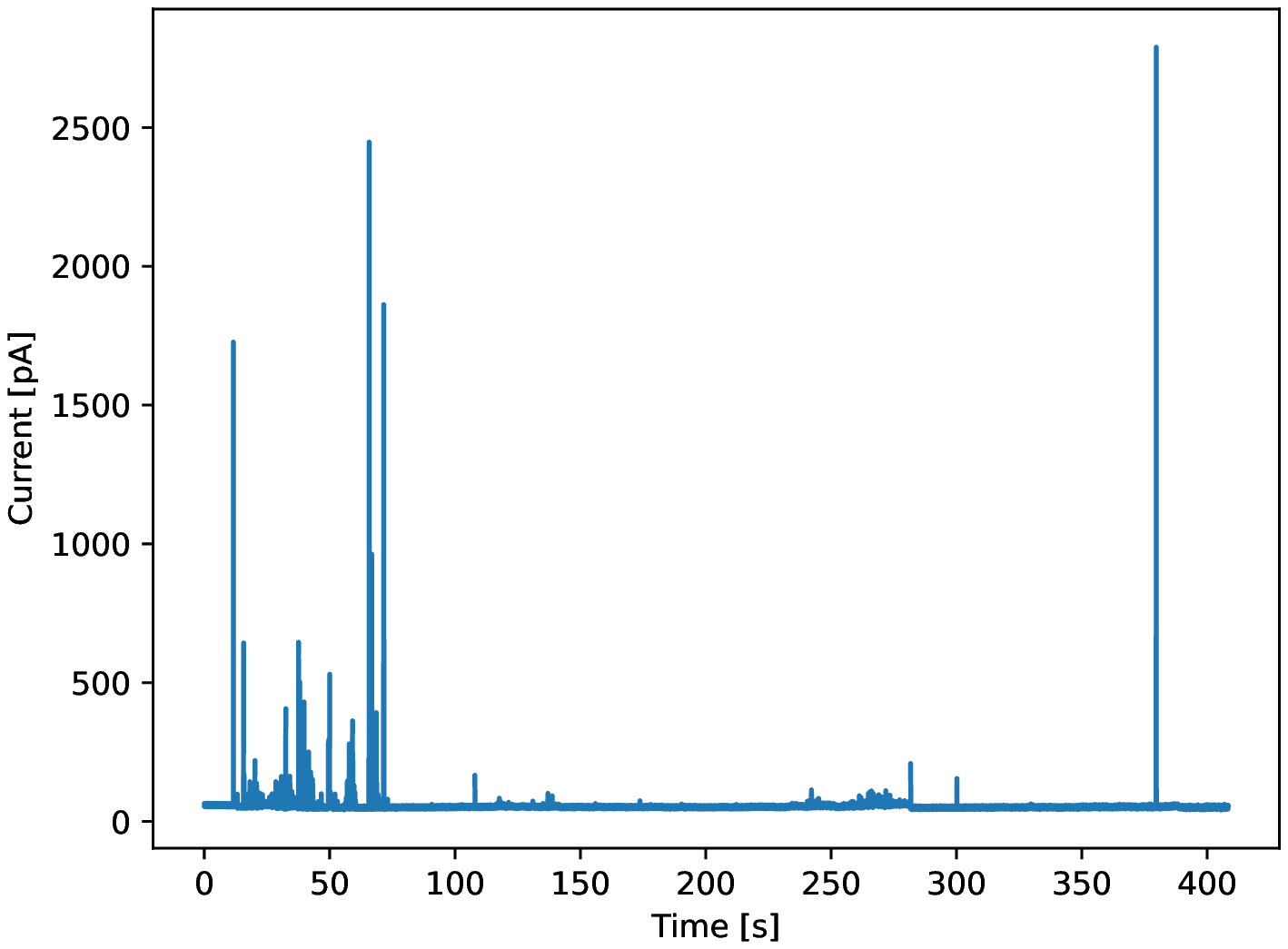}
\end{subfigure}\\
\begin{subfigure}[b]{0.4\textwidth}
\includegraphics[scale=0.4]{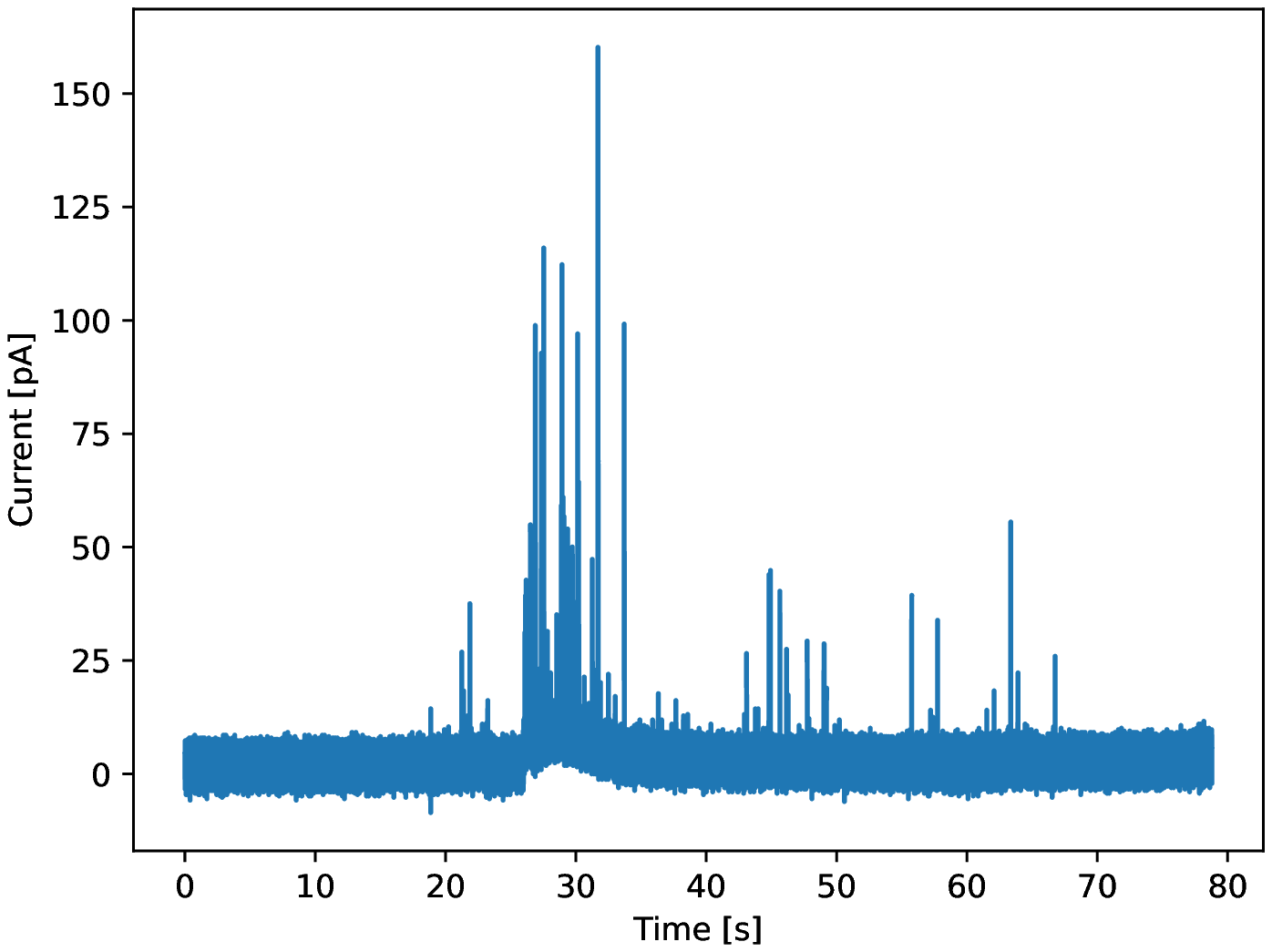}
\end{subfigure}
\begin{subfigure}[b]{0.4\textwidth}
\includegraphics[scale=0.4]{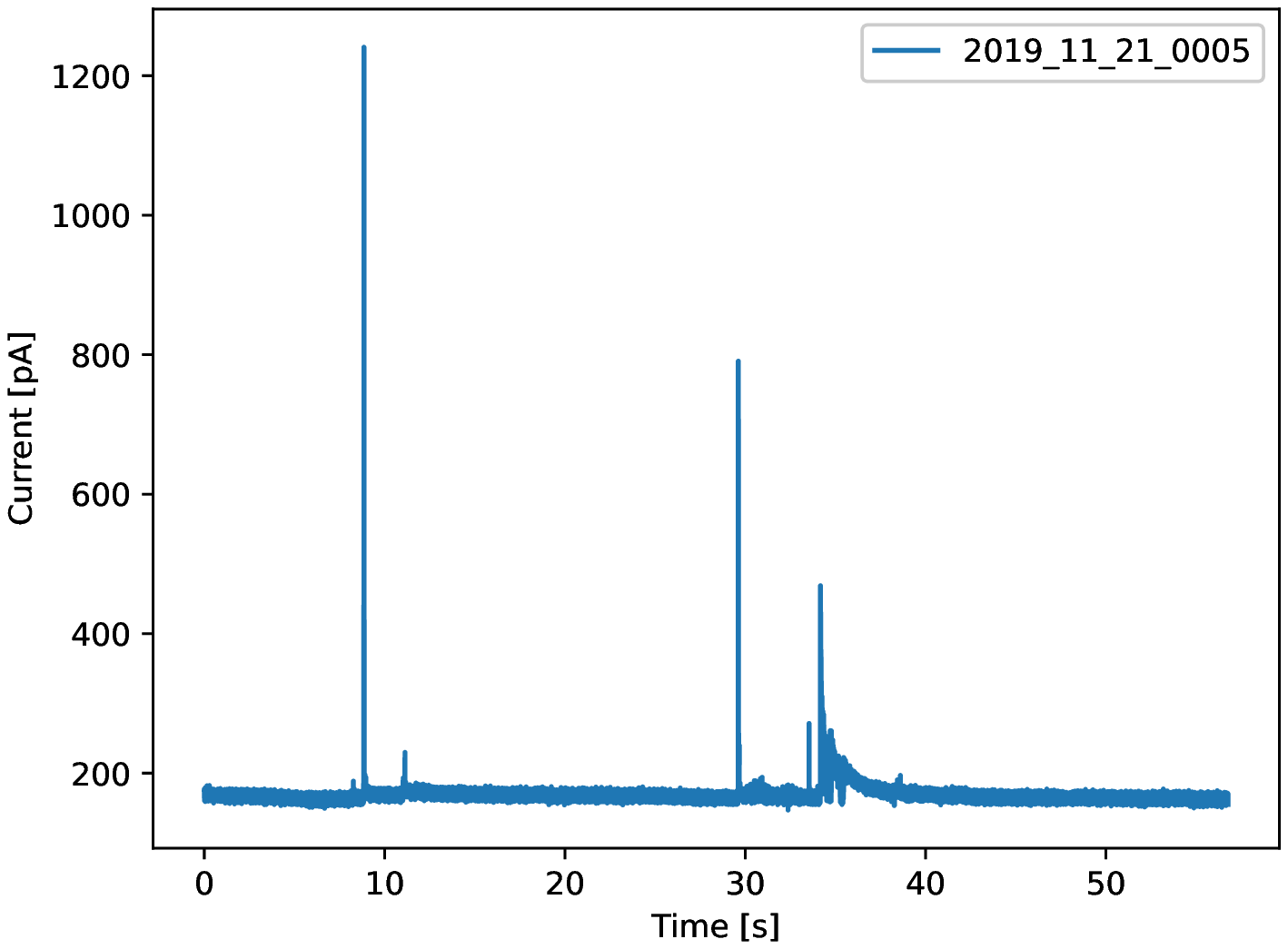}
\end{subfigure}
\caption{Amperometric traces of different cells (Chromaffin, \textit{Drosophila melanogaster}, PC12s cells, Beta cells).}
\end{figure}
%%%%%%%%%%%%%%%%%%%%%%%%%%%%%%%%%%%%%%%%%%%%%%%%%%%%%%%%%%%%%%%%%%%%%
\subsubsubsection*{PC12 cells}
PC12 cells were obtained as a gift from Lloyd Greene (Columbia University). The growth medium for PC12 cells consisted of $85$\% RPMI-1640 medium (Sigma-Aldrich, Sweden), $10$\% donor horseserum (Sigma-Aldrich, Sweden), and $5$\% fetal bovine serum (Sigma-Aldrich, Sweden). The cell culture was maintained at $37^\circ$C with an environment of $7$\% $\mathrm{CO}_2$ and $100$\% humidity. To achieve optimal growth condition, PC12 cells were cultured on commercial cell culture flasks pre-coated with type IV collagen (Corning BioCoat, Fisher Scientific, Sweden) and were propagated when confluence was reached, which was typically every $7$ days. For the experiments, PC12 cells were seeded on $60$ mm cell culture dishes pre-coated with type IV collagen (Corning BioCoat, Fisher Scientific, Sweden) $4-5$ days before the experiments.
%%%%%%%%%%%%%%%%%%%%%%%%%%%%%%%%%%%%%%%%%%%%%%%%%%%%%%%%%%%%%%%%%%%%%
\subsubsubsection*{Cell and Vesicle Isolation}
Bovine adrenal glands were obtained from a local slaughterhouse for isolating chromaffin cells. Briefly, the vein was perfused with Locke’s buffer to clear away blood cells. The medulla was isolated after collagenase ($0.2$\%, Roche, Sweden) treatment, and cells were isolated using a series of homogenization and centrifugation steps. For single cell experiments, $\approx 500000$ cells were seeded on collagen (IV) coated plastic dishes (Corning Biocoat, VWR, Sweden) and maintained in a humidified incubator at $37^\circ$ C, $5$\% $\mathrm{CO}_2$ for a maximum of $3$ days prior to experiments. 
For vesicle isolation, the medulla was mechanically homogenized in homogenizing buffer, and the vesicles were purified using a series of centrifugation steps: $1000$g for $10$ min to remove whole cells followed by $10000$g to pellet vesicles. All centrifugation was performed at $4^\circ$ C. The final pellet of vesicles was resuspended and diluted in homogenizing buffer and subsequently used for electrochemical measurements the same day.
%%%%%%%%%%%%%%%%%%%%%%%%%%%%%%%%%%%%%%%%%%%%%%%%%%%%%%%%%%%%%%%%%%%%%
\subsubsubsection*{Electrochemical recording for PC12 and Chromaffin cells
}
Electrochemical recordings from single PC12 and Chromaffin cells were performed on an inverted microscope (IX71/IX81, Olympus). For data collection, an Axopatch $200$B potentiostat instrument (Molecular Devices, Sunnyvale, CA) and a digital acquisition system (Digidata $1440$A, Molecular Devices) were used. A working electrode was held at $+700$ mV versus an Ag/AgCl reference electrode using the potentiostat. The signal output was acquired at $10$kHz, filtered at $2$ kHz using a low pass Bessel filter, and digitized at $5$ kHz.
For SCA, the working electrode was placed to top of a single cell to record electrochemical signals. Cells were stimulated for exocytosis for $5$ s and data were collected. While for IVIEC, the working electrode was pushed into a single cell to record signals.
%%%%%%%%%%%%%%%%%%%%%%%%%%%%%%%%%%%%%%%%%%%%%%%%%%%%%%%%%%%%%%%%%%%%%
\subsubsubsection*{Fly Dataset}
All flies were cultured on standard potato meal/agar medium at $25^\circ$ C. Drosophila lines were obtained from the Bloomington Stock Center (Bloomington IN, USA) and include: Tdc2-Gal4 and UAS-Chr2-mCherry. Larvae used for all experiments were F1 progeny of homozygous Tdc2-Gal4 and UAS-Chr2-mCherry. Prior to all experiments, larvae were fed $200$ mM all trans-retinal. By applying blue light, the conformation of ATR changes from a trans-conformation to cis-conformation and causes the cascade responsible for the exocytosis. This method of neuron-specific ChR2-mediate stimulation is specific and activates only one type neuron.

Third-instar larvae were dissected in HL$3.1$ saline that included $2.0$ mM calcium and $7$ mM L-glutamic acid to allow normal rates of exocytosis and to inhibit spontaneous muscle contractions. L-glutamic acid and calcium were added to HL$3.1$ on the day of experiment, and fresh stock solutions of L-glutamic acid were made weekly. Larvae were mounted on a Sylgard dissecting plate by placing two small pins at the posterior and interior end. A vertical cut was made along the dorsal midline toward the rostral end of the larva. The organs were removed and the fillet pinned laterally, slightly stretching the muscles of the body and generating a relatively flat surface. The pinned fillet was cleaned by rinsing with dissecting buffer $3$ times. The larval dissection and all the measurements were performed at room temperature. The fluorescent marker mCherry was visualized with an IX71S1F-2 Olympus microscope with $40$x objective and $620$ nm (maximum wavelength) filter set with a $150$ W xenon light source. 

A micromanipulator was used to position a $5-$micron carbon fiber disk microelectrode directly on to the Type II varicosities. An Ag/AgCl reference electrode was submerged in the HL$3.1$ bath solution. Stimulation of ChR2 was initiated by light emitted through the microscope objective ($40$x) by a $150$ W Xenon microscope lamp (MT20, Olympus, Japan) coupled to an emission filter ($437-456$ nm). All experiments were done in a dark room. Electrodes were held at an overpotential for oxidation of octopamine, 900 mVvs. an Ag/AgCl reference electrode (World Precision Instruments, Inc., Sarasota, FL), using a commercial patch-clamp instrument (Axopatch $200$B; Axon Instruments, Foster City, CA). The signal was digitized at $20$ kHz and filtered with an internal low pass $4-$pole Bessel filter at $2$ kHz. The signal was displayed in real time (AxoScope $8.1$; Axon Instruments) and stored digitally. 

% The collected data were analyzed with Igor Pro $6$ (Version $6.3.4.1$; WaveMetrics, Lake Oswego, OR) using an Igor Procedure. Data were considered significantly different at a $95$\% confidence level. Error bars are standard error of the mean (SEM). 

%%%%%%%%%%%%%%%%%%%%%%%%%%%%%%%%%%%%%%%%%%%%%%%%%%%%%%%%%%%%%%%%%%%%%
\subsubsubsection*{Beta cells}
For exocytosis experiments, the culture medium was removed and the beta cells were rinsed with Krebs buffer. During the analysis, beta cells were kept in the Krebs buffer solution at $37^\circ$C. Electrochemical recordings were performed on an inverted microscope, under the control of an Axopatch $200$B potentiostat (Molecular Devices, Sunnyvale, CA). To monitor exocytosis, the disk carbon fiber electrode was placed on the top of a beta cell by use of a Patch-Clamp Micromanipulator (PCS$-5000$, Burleigh Instruments, Inc., USA). High K+ concentration stimulating solution ($30$ mM KCl (Modified Krebs buffer solution), see Experimental Details – Reagents) was ejected onto the cells with a micropipette for $5$ s. The working potential of electrode was $+700$ mV vs a Ag/AgCl reference electrode and the output was filtered at $2.1$ kHz and digitized at $5$ kHz.
For intracellular analysis, chemical stimulation is not needed (Modified Krebs buffer solution). IVIEC was performed with the same instrument and the same buffer and temperature, as described above except on different beta cells in the same dish. The tip of the nano-tip carbon fiber electrode was first located on top of the single beta cell membrane and then, by use of a micromanipulator, the tip was gradually pressed through the membrane of the beta cell, while the current was recorded.
%%%%%%%%%%%%%%%%%%%%%%%%%%%%%%%%%%%%%%%%%%%%%%%%%%%%%%%%%%%%%%%%%%%%%
% \clearpage
\subsubsubsection*{VIEC Dataset}
\label{subsubsec:cells}

\begin{figure}[!htbp]
	\centering
	% \begin{subfigure}[b]{0.4\textwidth}
	% 	\includegraphics[scale=0.3]{IVIEC.png}
	% \end{subfigure}
	\begin{subfigure}[b]{0.4\textwidth}
		\includegraphics[scale=0.4]{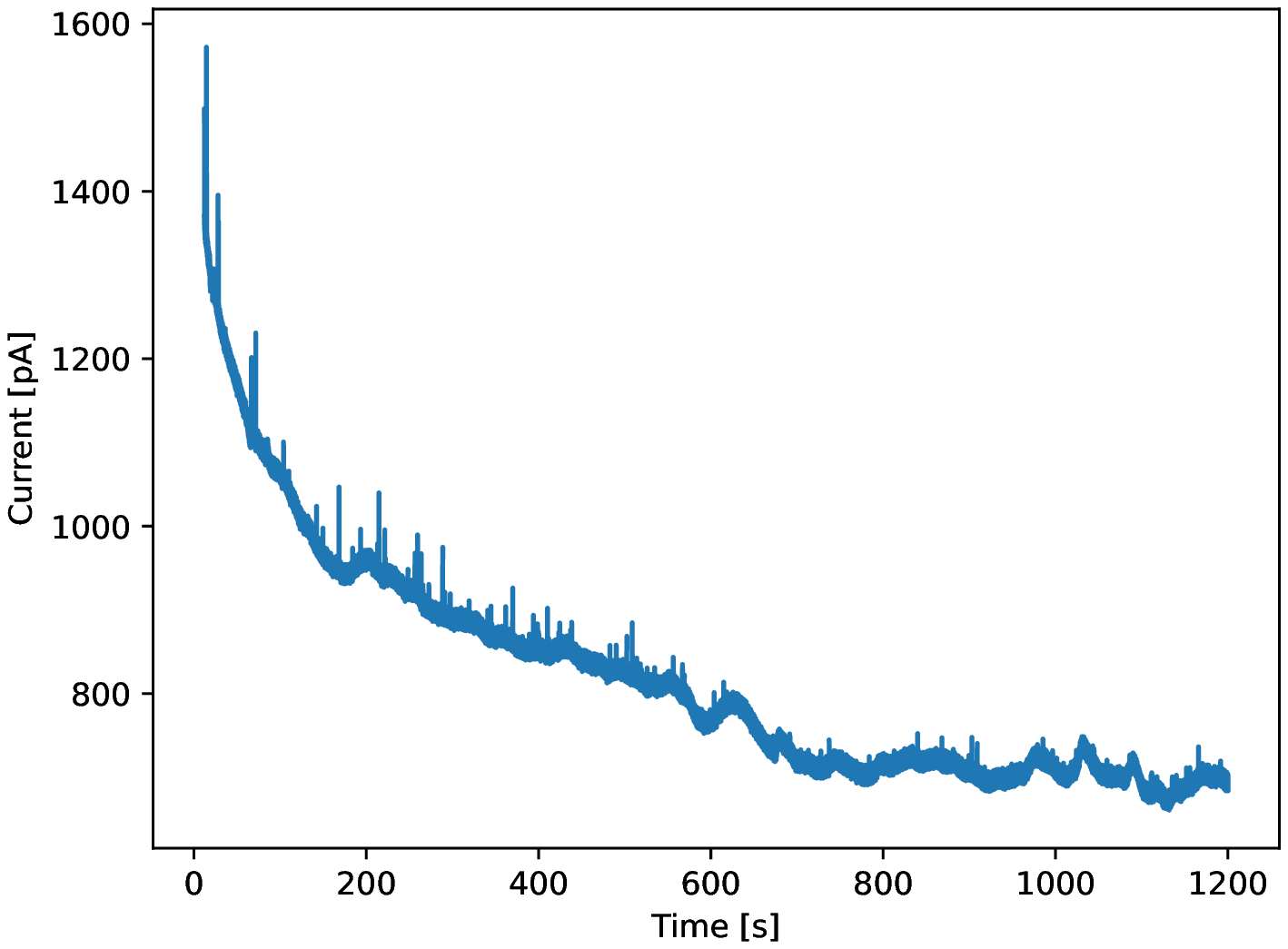}
	\end{subfigure}
	\begin{subfigure}[b]{0.4\textwidth}
		\includegraphics[scale=0.4]{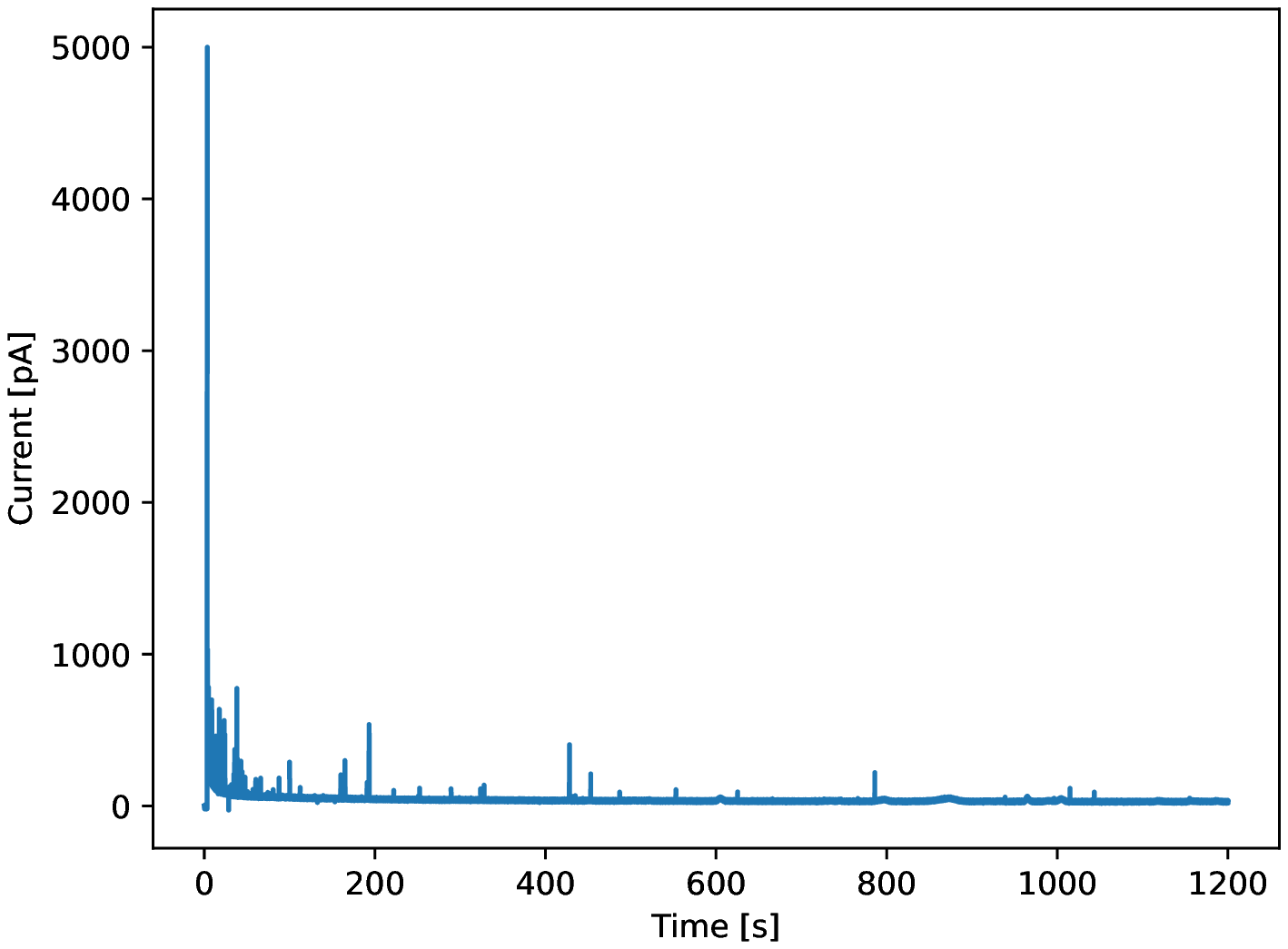}
	\end{subfigure}
	\caption{Amperometric traces of Chromaffin cells with different experimental approaches (VIEC - adding method, VIEC - dipping method). }
\end{figure}

The carbon fiber microdisk electrode was placed in the bath solution and connected to another amplifier (the working potential was set at $+900$ mV). The recorded RP and VIEC signals were filtered at $2$ kHz using a $4-$pole Bessel filter and digitized at $10$ kHz using a Digidata model $1440$A with Axoscope $10.3$ software.

After subtracting the baseline of a VIEC raw trace, the VIEC current spikes were searched and counted in each cycle. The detection limit used was $6$ times the standard deviation of the baseline noise in a whole cycle. The area of each spike was the total charge transferred from the released catecholamine to the electrode. The vesicular content in moles was obtained by use of Faraday’s Law, in which the charge transfer number of electrons is $2$. All traces were visually inspected, and false positives were removed. 
%%%%%%%%%%%%%%%%%%%%%%%%%%%%%%%%%%%%%%%%%%%%%%%%%%%%%%%%%%%%%%%%%%%%%
\clearpage
\subsection{Setting up \tsf\ on cluster}
\label{subsec:tsfresh_setup}
% environment variables

When handling large datasets such as the \hsd\ dataset, VIEC dataset and the electrode Dataset, data manipulations on a PC will frequently lead to memory overflow even with chunk sizes as small as $\sim$1000. Therefore, to calculate accuracy on large chunk sizes, it is recommended to work on a server/cluster. 
For using \tsf\ on a server, please follow the instructions on the \tsf\ official website.

Note depending on the \textit{numpy} and \textit{pandas} version activated in the $python$ environment, \tsf\ may report various error messages. For the sake of simplicity, it is recommended to set up a virtual environment for version management. The combination that we found running bug-free and had no dependency issues is as follows:
\begin{itemize}
	\item[] {\makebox[4cm]{\textbf{Package}:\hfill} \textbf{Version}}
	\item[] {\makebox[4cm]{\textit{Python}:\hfill} 3.6.8}
    \item[] {\makebox[4cm]{\textit{numpy}:\hfill} 1.19.1}
    \item[] {\makebox[4cm]{\textit{pandas}:\hfill} 0.23.4}
    \item[] {\makebox[4cm]{\textit{scikit-learn}:\hfill} 0.23.1}
    \item[] {\makebox[4cm]{\textit{tsfresh}: \hfill} 0.16.0}
\end{itemize}
%%%%%%%%%%%%%%%%%%%%%%%%%%%%%%%%%%%%%%%%%%%%%%%%%%%%%%%%%%%%%%%%%%%%%%%%%%%
\subsection{\laz}
\label{subsec:lazy}

Once the \tsf\ features are extracted, one could evaluate different established ML classification models using \laz\ even on the local PC, as the memory requirement is much lower than \tsf. For the installation of \laz\ on both PC and cluster, please refer to the tutorial in \url{https://lazypredict.readthedocs.io/en/latest/installation.html}. Mostly, \laz\ works in combination with the $scikit-learn$ ($sklearn$) library for normalization, splitting the train/test, and other data manipulations. For detailed information, see the linked website above.  
%%%%%%%%%%%%%%%%%%%%%%%%%%%%%%%%%%%%%%%%%%%%%%%%%%%%%%%%%%%%%%%%%%%%%%%%%%%
\subsection{Description of Top Features}
\label{subsec:features}

As discussed above, we observed $\approx 1\%$ features with relative importance value above $0.01$ for the all datasets. Large parts of the features description that follows here has been adapted from \tsf\ documentation \url{https://tsfresh.readthedocs.io/en/latest/text/_generated/tsfresh.feature_extraction.feature_calculators.html}.
%%%%%%%%%%%%%%%%%%%%%%%%%%%%%%%%%%%%%%%%%%%%%%%%%%%%%%%%%%%%%%%%%%%%%%%%%%%
\subsubsection*{The Top Most Features in Each Dataset by Highest Importance Value}
\label{subsubsec:top}

The feature with the highest importance value for each of the candidate dataset is discussed here.
%%%%%%%%%%%%%%%%%%%%%%%%%%%%%%%%%%%%%%%%%%%%%%%%%%%%%%%%%%%%%%%%%%%%%%%%%%%
\subsubsubsection*{number\_peaks}
% https://tsfresh.readthedocs.io/en/latest/api/tsfresh.feature_extraction.html#tsfresh.feature_extraction.feature_calculators.number_peaks
The \tsf\ function \textit{number\_peaks(x,n)} calculates the number of peaks of at least support n in the time series x. A peak of support n is defined as a subsequence of x where a value occurs, which is bigger than its n neighbors to the left and to the right. Hence in the sequence x = [$3, 0, 0, 4, 0, 0, 13$], 4 is a peak of support 1 and 2 because in the subsequences [$0, 4, 0$] and [$0, 0, 4, 0, 0$], 4 is still the highest value. However, 4 is not a peak of support 3 because 13 is the $3^{\mathrm{rd}}$ neighbour to the right of 4 and is greater than 4.

\subsubsubsection*{spike\_welch\_density (or) spkt\_welch\_density(x, param)}
% https://tsfresh.readthedocs.io/en/latest/api/tsfresh.feature_extraction.html#tsfresh.feature_extraction.feature_calculators.spkt_welch_density
The \tsf\ function \textit{spkt\_welch\_density(x, param)} returns the power spectrum of the different frequencies given the time series $x$, and param that contains dictionaries with coefficients $x$, where $x$ is an integer.

\subsubsubsection*{permutation\_entropy\_dimension (or) permutation\_entropy(x, tau, dimension)}
% https://tsfresh.readthedocs.io/en/latest/api/tsfresh.feature_extraction.html#tsfresh.feature_extraction.feature_calculators.permutation_entropy
The \tsf\ function \textit{permutation\_entropy(x, tau, dimension)} chunks the data into sub-windows for every $\tau$; replaces each D-window by the permutation that captures the ordinal ranking of the data; then counts the frequencies of every permutation and returns their entropy. For example, given a vector $x = [4, 7, 9, 10, 6, 11, 3]$ with $D=3$ and $\tau = 1$ is turned into $[[4, 7, 9][ 7, 9, 10], [ 9, 10, 6], [10, 6, 11], [ 6, 11, 3]]$. Each D-window is replaced by the permutation that gives, $[[0, 1, 2], [1, 2, 0], [1, 0, 2], [1, 2, 0]]]$

\subsubsubsection*{fft\_aggregated\_\enquote{kurtosis} (or) feature\_calculators.fft\_aggregated(x, param)}
% https://tsfresh.readthedocs.io/en/latest/api/tsfresh.feature_extraction.html#tsfresh.feature_extraction.feature_calculators.fft_aggregated
This \tsf\ function returns kurtosis of the absolute fourier transform spectrum given a time series, $x$ and param list that contains including kurtosis values.
%%%%%%%%%%%%%%%%%%%%%%%%%%%%%%%%%%%%%%%%%%%%%%%%%%%%%%%%%%%%%%%%%%%%%%%%%%%
\subsubsection*{Simple Statistical Features Common Across Different Datasets}

\subsubsubsection*{median or median(x)}
% https://tsfresh.readthedocs.io/en/latest/api/tsfresh.feature_extraction.html#tsfresh.feature_extraction.feature_calculators.median
This \tsf\ function returns the median of input time series $x$.

\subsubsubsection*{mean or mean(x)}
% https://tsfresh.readthedocs.io/en/latest/api/tsfresh.feature_extraction.html#tsfresh.feature_extraction.feature_calculators.mean
This \tsf\ function returns the mean of input time series $x$.

\subsubsubsection*{minimum or minimum(x)}
% https://tsfresh.readthedocs.io/en/latest/api/tsfresh.feature_extraction.html#tsfresh.feature_extraction.feature_calculators.minimum
This \tsf\ function returns the lowest value of input time series $x$.

\subsubsubsection*{quantile (or) value\_quantile\_q\_{x}  with q = $0.1, 0.2, 0.3, 0.4, 0.5(\mathrm{median}), 0.6, 0.7, 0.8$}
% https://tsfresh.readthedocs.io/en/latest/api/tsfresh.feature_extraction.html#tsfresh.feature_extraction.feature_calculators.quantile
This \tsf\ function calculates the q-th quantile of x, \ie\ the value of x greater than $q \times 100\%$ of the ordered values from $x$. A special quantile is $q=0.5$, which coincides with the median of a time series $x$. 
%%%%%%%%%%%%%%%%%%%%%%%%%%%%%%%%%%%%%%%%%%%%%%%%%%%%%%%%%%%%%%%%%%%%%%%%%%%
\subsubsection{Common Features Across All Datasets}

\subsubsubsection*{c3\_lag\_3 (or) c3(x, lag)}
% https://tsfresh.readthedocs.io/en/latest/api/tsfresh.feature_extraction.html#tsfresh.feature_extraction.feature_calculators.c3
This \tsf\ function uses c3 statistics to measure non-linearity in the time series given a time series $x$ and a lag parameter that should be used in the calculation of the feature.

Another common feature that emerged across all datasets is the quantile which is already discussed in the previous subsection.
%%%%%%%%%%%%%%%%%%%%%%%%%%%%%%%%%%%%%%%%%%%%%%%%%%%%%%%%%%%%%%%%%%%%%%%%%%%
\subsubsection{Common Features Across All Methods in the \hsd\ Dataset}

\subsubsubsection*{mean\_absolute\_change (or) value\_mean\_abs\_change}
% https://tsfresh.readthedocs.io/en/latest/api/tsfresh.feature_extraction.html#tsfresh.feature_extraction.feature_calculators.mean_abs_change
This \tsf\ function returns the mean over the absolute differences between subsequent time series values which is
$$\frac{1}{n} \sum_{i=1}^{n-1} |x_{i+1} - x_{i}|,$$
that evaluates the total variation of a given time series $\{x\}$.

\subsubsubsection*{ar\_coefficient (or) value\_ar\_coefficient\_coeff\_{i}\_k\_10, i=$2,3,4,5$}
% https://tsfresh.readthedocs.io/en/latest/api/tsfresh.feature_extraction.html#tsfresh.feature_extraction.feature_calculators.ar_coefficient
The \tsf\ function \textit{ar\_coefficient(x,param)} fits the unconditional maximum likelihood
of an autoregressive AR(k) process: a stochastic process to predict future behavior based on past behavior, where k parameter is the maximum lag of the process (history length). The autoregression can mathematically be expressed as 
$$ X_{t}=\varphi_0 +\sum _{{i=1}}^{k}\varphi_{i}X_{{t-i}}+\varepsilon_{t},$$
where $X_{{t-i}}$ represent the past series values in \hsd, $\varepsilon_{t}$ represents some white noise and $\varphi_{i}$ denote the autoregression coefficients, here $i=2,3,4,5$ are found to dominate the classification process. 

\subsubsubsection*{change\_quantile (or) value\_change\_quantiles\_f\_agg \\ \_"mean"\_isabs\_True\_qh\_1.0\_ql\_0.0}
The \tsf\ function \textit{change\_quantiles(x, ql, qh, isabs, f\_agg)} first fixes a corridor given by the lowever quantile ql and the higher quantile qh of the \textbf{distribution} of x. Then it calculates the average(f\_agg = "mean"), absolute value(isabs=True) of consecutive changes of the series x inside this corridor. Since the ql and qh we are looking at are $0.0$ and $1.0$, respectively, the resulting function value will be exactly identical to the previous feature, \textbf{value\_mean\_abs\_change}.

Couple of other features that were common across all methods in the \hsd\ dataset were number of peaks and median, which has already been discussed in the previous subsections. Further, similar features were observed when increasing the amount of forests, and the similarity grows correspondingly. 
%%%%%%%%%%%%%%%%%%%%%%%%%%%%%%%%%%%%%%%%%%%%%%%%%%%%%%%%%%%%%%%%%%%%%%%%%%%%%%%%%%%%%%%%%%%%%%%%%%%%%
\subsubsection{Common Features Across All Methods in the Electrodes Dataset}

Both permutation\_entropy and minimum have been discussed in the previous subsections.
%%%%%%%%%%%%%%%%%%%%%%%%%%%%%%%%%%%%%%%%%%%%%%%%%%%%%%%%%%%%%%%%%%%%%%%%%%%%%%%%%%%%%%%%%%%%%%%%%%%%%
\subsubsection{Common Features Across All Methods in the Cells Dataset}

Both permutation\_entropy and quantiles have been discussed in the previous subsections.
%%%%%%%%%%%%%%%%%%%%%%%%%%%%%%%%%%%%%%%%%%%%%%%%%%%%%%%%%%%%%%%%%%%%%%%%%%%%%%%%%%%%%%%%%%%%%%%%%%%%%
\subsubsection{Common Features Across All Methods in VIEC Dataset}

fft\_aggregated\_\enquote{kurtosis} has been discussed in the previous subsections.
%%%%%%%%%%%%%%%%%%%%%%%%%%%%%%%%%%%%%%%%%%%%%%%%%%%%%%%%%%%%%%%%%%%%%%%%%%%%%%%%%%%%%%%%%%%%%%%%%%%%%%%%%%%%%%%%%%%%%%%%%%%%%%%%%%%%%%%%%%%%%%%%%%%%%%%%%%%%%%%%%%%%%%%%%%%%%%%%%%%%%%%%%%%%%%%%%%%%%%%%%%
\clearpage
\subsection{Code}
\label{subsec:code}

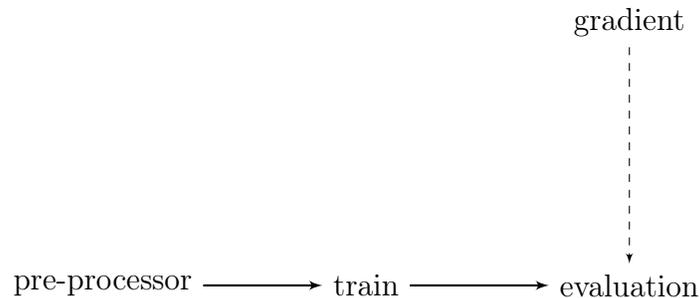
\begin{figure}[!htbp]
\centering
\begin{tikzpicture}[node distance = 3.5cm, auto]
  \tikzstyle{decision} = [diamond, draw, fill=blue!20,
    text width=5em, text badly centered, node distance=1cm, inner sep=0pt]
  \tikzstyle{block} = [rectangle, draw,
    text width=5.8em, text centered, rounded corners, minimum height=2em]
  \tikzstyle{line} = [draw, -latex']  
  \tikzstyle{cloud} = [draw, ellipse,node distance=3cm,
    minimum height=1em, text width=5em]
    
    \node[] at (0,0) (pre) {pre-processor};
    \node[right of=raw] (train) {train};
    \node[right of=train] (main) {evaluation};
    \node[above of=main] (grad) {gradient};

    \path [line, thick] (pre) -- (train);
    \path [line, thick] (train) -- (main);
    \path [line, dashed] (grad) -- (main);

\end{tikzpicture}
\caption{Structure of the Decision Tree Analysis Package. All programs fetch parameters set by the use in the \texttt{params} module.}
\label{fig:package}
\end{figure}

The classification scripts need to be executed in the order of arrows in the illustration shown in \fig\ \ref{fig:package}. The \texttt{params} module defines the global settings for classification. Most important parameters are: data path, chunk size, number of forests and number of decision trees per forest. The \texttt{gradient\_visualizer} module provides a visualization of the raw time series data. The path name should be adapted here. 
The \texttt{pre-processor} module splits the raw data into chunks and assigns each chunk with a label. The output files are stored in \texttt{./dataframe/}. The \texttt{train} module extracts \textit{\tsf} features from each chunk and the output files are stored in \texttt{./features/}. This script requires high computational power and is therefore recommended to run on the server. The \texttt{evaluation} module trains the decision tree classifier (including random forest, extra trees and XGBoost) and returns the final output: \begin{enumerate*} \item mean and standard deviation of accuracy \item features with importance values \item heat map of cross-correlation between features \item bar plot of the feature importances \item visualization of one of the decision trees \end{enumerate*}

Training a \rf\ with a small sample size can be very efficient on a PC, however, when it comes to the classification task with $7000$ data samples as in the Hofmeister Series case, training on the cluster significantly outperforms a PC (scaling factor $\sim$20). 
The \texttt{decision-tree} package is implemented in Python3.2.0 and the most recent version of the implementation of \rf\ training and evaluation can be found in our $github$ repository: \href{https://github.com/krishnanj/TBLAD}{https://github.com/krishnanj/TBLAD}.
%%%%%%%%%%%%%%%%%%%%%%%%%%%%%%%%%%%%%%%%%%%%%%%%%%%%%%%%%%%%%%%%%%%%%%%%%%%%%%%%%%%%%%%%%%%%%%%%%%%%%
%%%%%%%%%%%%%%%%%%%%%%%%%%%%%%%%%%%%%%%%%%%%%%%%%%%%%%%%%%%%%%%%%%%%%%%%%%%%%%%%%%%%%%%%%%%%%%%%%%%%%
\end{suppinfo}
%%%%%%%%%%%%%%%%%%%%%%%%%%%%%%%%%%%%%%%%%%%%%%%%%%%%%%%%%%%%%%%%%%%%%
%%%%%%%%%%%%%%%%%%%%%%%%%%%%%%%%%%%%%%%%%%%%%%%%%%%%%%%%%%%%%%%%%%%%%
%% The appropriate \bibliography command should be placed here.
%% Notice that the class file automatically sets \bibliographystyle
%% and also names the section correctly.
%%%%%%%%%%%%%%%%%%%%%%%%%%%%%%%%%%%%%%%%%%%%%%%%%%%%%%%%%%%%%%%%%%%%%
\clearpage
\bibliography{references}
%%%%%%%%%%%%%%%%%%%%%%%%%%%%%%%%%%%%%%%%%%%%%%%%%%%%%%%%%%%%%%%%%%%%%
\end{document}